\documentclass[11pt]{article}
\pdfoutput=1
\usepackage[makeroom]{cancel}
\usepackage{cite}
\usepackage{graphicx}
\usepackage{subfigure}
\usepackage{multirow} 
\usepackage[utf8]{inputenc}
\usepackage[english]{babel}
\usepackage{graphicx}
\usepackage{amsmath}
\usepackage{multicol,lipsum,graphicx,float}
\usepackage{color}
\usepackage{cite}
\usepackage{blindtext}
\usepackage{hyperref}
\usepackage{amsmath,amssymb}
\usepackage[bottom]{footmisc}
\newcommand{\be}{\begin{equation}}
\newcommand{\ee}{\end{equation}}
\newcommand{\bea}{\begin{eqnarray}}
\newcommand{\eea}{\end{eqnarray}}
\makeatletter

\makeatletter
\newcommand*{\rom}[1]{\expandafter\@slowromancap\romannumeral #1@}
\makeatother

\begin{document}
 \begin{center}
 {\Large\bf Scalar assisted singlet doublet fermion dark matter model and 
electroweak vacuum stability}
 \\
 \vskip .5cm
 {
 Amit Dutta Banik$^{a,}$\footnote{amitdbanik@iitg.ac.in},
 Abhijit Kumar Saha$^{a,}$\footnote{abhijit.saha@iitg.ac.in},
 Arunansu Sil$^{a,}$\footnote{asil@iitg.ac.in}}\\[3mm]
 {\it{
 $^a$ Department of Physics, Indian Institute of Technology Guwahati, 781039 Assam, India}
 }
 \end{center}
 \vskip .5cm
 \begin{abstract}
We extend the so-called singlet doublet dark matter model, where the dark matter is an admixture of a 
Standard Model singlet and a pair of electroweak doublet fermions, by a singlet scalar field.  
The new portal coupling of it with the dark sector not only contributes to the dark matter 
phenomenology (involving relic density and direct detection limits), but also becomes important for 
generation of dark matter mass through its vacuum expectation value. While the presence of dark sector
fermions affects the stability of the electroweak vacuum adversely, we find this additional singlet is
capable of making the electroweak vacuum absolutely stable upto the Planck scale. A combined study of
dark matter phenomenology and Higgs vacuum stability issue reflects that the scalar sector mixing angle
can be significantly constrained in this scenario.

%
 \end{abstract}

\section{Introduction}
Although the discovery of the 125 GeV Higgs boson at Large Hadron Collider
(LHC) \cite{Aad:2014aba, Chatrchyan:2012ft} undoubtedly 
marks the ultimate success of the Standard Model (SM), there are issues in particle physics and cosmology, 
supported by observations, which can not be explained in the SM framework. For example,  SM justifies 
only 5\% of the total matter content of the Universe preferably known as visible matter. Compelling  
evidences from astrophysical and cosmological observations of cosmic microwave background radiation 
(CMBR), spiral galaxy rotation curve, colliding clusters etc. indicate the presence of unknown matter, called 
dark matter (DM) which constitutes 25\% of the Universe. There are some other theoretical issues for which 
SM can not provide clear answer. In particular, it is well known that the Higgs quartic coupling ($\lambda_H$) 
turns negative at energy scale $\Lambda_I^{\textrm{SM}}\sim 10 ^{10}$ GeV \cite{Buttazzo:2013uya,
Degrassi:2012ry,Tang:2013bz,Ellis:2009tp,EliasMiro:2011aa} (with $m_t=173.2$ GeV \cite{Khachatryan:2015hba}) leading to a possible instability of the electroweak 
(EW) minimum\cite{Isidori:2001bm}. However the conclusion crucially depends on precise value of the top quark and Higgs mass. In 
presence of a deeper minimum compared to the EW one, question will also arise why the Universe has chosen 
the EW vacuum over the global minimum \cite{Espinosa:2007qp,Lebedev:2012sy,Hook:2014uia,Kobakhidze:2013tn,Fairbairn:2014zia,Kearney:2015vba}. 

In order to circumvent these shortcomings of the SM, one has to introduce new physics beyond the 
Standard Model. In an earlier attempt \cite{Ghosh:2017fmr}, the SM is extended with two SM singlet scalars,
one is with zero and other has non-zero vacuum expectation value (vev).  
It is shown in \cite{Ghosh:2017fmr} that while the singlet scalar with zero vev plays the role of the DM, the other scalar 
with non-zero vev mixes with SM Higgs (Higgs portal) and affect the dark matter phenomenology in such a way that the scalar
DM having mass $\sim 200$ GeV and onward can satisfy the relic density and direct search constraints 
from LUX\cite{Akerib:2016vxi}, XENON-1T\cite{Aprile:2017iyp}, Panda 2018\cite{Cui:2017nnn} 
and XENON-nT\cite{Aprile:2015uzo}. On the other hand, it turns out that the interaction of the 
scalar fields with SM Higgs can modify the instability scale ($\Lambda_I$) to a value larger than $\Lambda_I^{\textrm{SM}}$ 
by several order of magnitude. In fact the scalar with non-zero vacuum expectation value 
having mass smaller than $\Lambda_I^{\textrm{SM}}$ can indeed make the electroweak 
vacuum absolutely stable \cite{Ghosh:2017fmr} with 
the help of threshold effect\cite{Gogoladze:2008gf,EliasMiro:2012ay,Lebedev:2012zw}.
Other works involving DM and EW vacuum stability
can be found in \cite{Khan:2014kba,Khoze:2014xha,Gonderinger:2009jp,Garg:2017iva,Chen:2012faa,Costa:2014qga,Oda:2017kwl,
Chakrabarty:2014aya,Chakrabarty:2015yia,Chakrabarty:2016smc}.
The extra scalar field(s) could also be connected to several other unresolved physics of the Universe involving   
inflation\cite{Bhattacharya:2014gva,Saha:2016ozn,Oda:2017zul} or 
neutrinos\cite{Datta:2013mta,Coriano:2014mpa,Ng:2015eia,Bonilla:2015kna,Haba:2015rha,Chakrabortty:2013zja,Khan:2012zw} etc.

In this work we consider the singlet doublet dark matter (SDDM) scenario and explore how it can be extended minimally 
(if required) so as to achieve the 
EW vacuum stability till $M_P$. In a typical SDDM model \cite{Cai:2016sjz,Mahbubani:2005pt,DEramo:2007anh,
Enberg:2007rp,Cohen:2011ec,Cheung:2013dua,Calibbi:2015nha,Horiuchi:2016tqw,Banerjee:2016hsk
,Abe:2017glm,Xiang:2017yfs,Maru:2017otg,Maru:2017pwl,Calibbi:2018fqf,Esch:2018ccs,Arcadi:2018pfo}, the dark sector is made up with two Weyl fermion doublets 
and one Weyl singlet fermion.
The Yukawa interactions of them with SM Higgs result three neutral fermion states, the lightest of which 
becomes a viable candidate for DM provided the stability is guaranteed by some symmetry argument. Unlike 
Higgs portal dark matter models, singlet doublet dark matter scenario directly couples the mass and 
dynamics of dark sector with the SM gauge sector. This is analogous to the case of supersymmetric 
extensions\cite{ArkaniHamed:2006mb} where the supersymmetry breaking scale provides mass of dark matter \cite{Calibbi:2015nha}.
Singlet doublet dark matter models also induce considerable co-annihilation effect which is absent in usual 
Higgs portal DM scenarios. Another interesting feature of SDDM model is related to evading the direct 
detection bound with some specified  ``blind spots''  of the model \cite{Calibbi:2015nha}. 

The SDDM carries different phenomenology from the usual extension of dark sector with vector-like fermion 
doublet and singlet \cite{Bhattacharya:2016lts,Bhattacharya:2016rqj,Bhattacharya:2017sml,Bhattacharya:2015qpa,Yaguna:2015mva,Narendra:2017uxl} due to the involvement of three 
neutral Majorana fermions in SDDM as compared to two vector like neutral 
fermions in \cite{Bhattacharya:2016lts,Bhattacharya:2016rqj,Bhattacharya:2017sml,Bhattacharya:2015qpa,Yaguna:2015mva}. In case of vector 
like singlet doublet models, it is possible to have interaction with $Z$ boson which can enhance the spin independent 
dark matter nucleon cross-section considerably. On the other hand, in case of SDDM, such interaction is 
suppressed \cite{Calibbi:2015nha}. Although in SDDM, spin dependent interaction (i.e. axial vector interaction) survives, 
the bounds on spin dependent dark matter nucleon cross-section\cite{Fu:2016ega} is not that stringent compared to spin 
independent limits and hence remain well below the projected upper limits. Therefore it relaxes the 
bounds on model parameters in the singlet doublet model allowing the model to encompass a large range 
of parameter space. 

Although the SDDM has many promising features as mentioned above, it also 
has some serious issues with the Higgs vacuum stability. The model involves new fermions, which can affect 
the running of Higgs quartic coupling leading to instability at high energy scale \cite{Egana-Ugrinovic:2017jib}. In an attempt to 
solve the Higgs vacuum stability  where the DM is part of the 
 SDDM model, we propose an extension of the SDDM with a  SM singlet scalar. 
We employ a $Z_4$ symmetry under which all the beyond SM fields carry non-trivial charges while SM fields 
are not transforming. The salient features of our model are the followings: 
\begin{itemize}

\item There exists a coupling between the additional scalar and the singlet Weyl fermion which eventually contributes to 
the mass matrix involving three neutral Weyl fermions. After the SM Higgs doublet and the scalar
get vevs,  mixing between neutral singlet fermion and doublet Weyl fermions occur and the lightest neutral 
fermion can serve as a stable Majorana dark matter protected by the residual $Z_2$ symmetry.  In this 
way, the vev of the additional scalar contributes to the mass of the DM as well as the mixing.

\item Due to the mixing between this new scalar 
and the SM Higgs doublet, two physical Higgses will result in this set-up. One of these would be
identified with the Higgs discovered at LHC. This  set-up therefore introduces a rich DM phenomenology (and different as compared
to usual SDDM model) as 
the second Higgs would also contribute to DM annihilation and the direct
detection cross-section.

\item The presence of the singlet scalar with non-zero vev helps in achieving the absolute stability of 
the EW vacuum. Here the mixing between singlet doublet scalars (we call it scalar mixing) plays an important
role. Hence the combined analysis of DM phenomenology (where this scalar mixing also participates) and vacuum
stability results in constraining this scalar mixing at a level which is even stronger than the existing limits on it
from experiments.
  \end{itemize}

The paper is organized as follows. In Sec.~\ref{model}, we describe the singlet scalar extended SDDM model.
Various theoretical and observational limits on the specified model are 
presented in Sec.~\ref{constraints}. 
 In the next section, we present our strategy, the related expressions including Feynman diagrams for studying
 dark matter phenomenology of this model. The discussion on the allowed 
parameter space of the model in terms of satisfying the DM relic density and 
direct detection limits are also mentioned in this Sec.~\ref{dm}.
In Sec.~\ref{vs}, the strategy to achieve vacuum stability of the scalar enhanced singlet doublet model is presented. 
In Sec.~\ref{EWph}, we elaborate on how to constrain parameters of the model while having a successful DM
candidate with absolute vacuum stability within the framework. Finally
the work is concluded with conclusive remarks in Sec.~\ref{conclusion}.

\section{The Model}
\label{model}

Like the usual singlet doublet dark matter model\cite{Mahbubani:2005pt,DEramo:2007anh,Enberg:2007rp,
Cohen:2011ec}, here also we extend the SM  framework 
by introducing two doublet Weyl fermions, $\psi_{D_1}, \psi_{D_2}$ and a singlet Weyl fermion 
field $\psi_S$. The doublets are carrying equal and opposite hypercharges ($Y=\frac{1}{2}(-\frac{1}{2})$ 
for $\psi_{{D_1}(D_2)}$) as required from gauge anomaly cancellation. Additionally the scalar 
sector is extended by including a SM real singlet scalar field, $\phi$. There exists a $Z_4$ symmetry, 
under which only these additional fields are charged which are tabulated in Table~\ref{tab1}. 
The purpose of introducing this $Z_4$ is two fold: firstly it avoids a bare mass term for the $\psi_S$ field. 
Secondly, although the $Z_4$ is broken by the vev of the $\phi$ field, there prevails a residual 
$Z_2$ under which all the extra fermions are odd. Hence the lightest combination of them is 
essentially stable. Note that with this construction, not only the DM mass involves vev of $\phi$ 
but also the dark matter phenomenology becomes rich due to the involvement of two physical 
Higgs (as a result of mixing between $\phi$ and the SM Higgs doublet $H$). Apart from these, 
$\phi$ is also playing a crucial role in achieving electroweak vacuum stability. The purpose of $\phi$ will 
be unfolded as we proceed. For the moment, we split our discussion into two parts first as extended  
fermion and next as scalar sectors of the model. 


\begin{table}
\begin{center}
\vskip 0.5 cm
\begin{tabular}{|c|c|c|c|c|}
\hline
Symmetry       & $\psi_{D_1}$ &   $\psi_{D_2}$ & $\psi_S$ & $\phi$  \\
\hline
$Z_4$         &    -i     &      i     &   i    &    -1    \\
\hline
\end{tabular}
\end{center}
\caption{Particle multiplets and their transformation properties under $Z_4$ symmetry.}
\label{tab1}
\end{table}

\subsection{Extended fermion sector}
\label{fermion}

The dark sector fermions $\psi_{D_1}, \psi_{D_2}$ and $\psi_S$ are represented as,
\be
\psi_{D_1} = \left ( \begin{array}{c} \psi_1^0 \\ \psi_1^- \end{array} \right ) ~~:[2, \frac{1}{2}]\,\,,~~~~~
\psi_{D_2} = \left ( \begin{array}{c} \psi_2^+ \\ \psi_2^0 \end{array} \right )~~:[2, -\frac{1}{2}]\,\,,~~~~~
\psi_S ~~:[1, 0]\,\, .
\label{eq1}
\ee
Here field transformation properties under SM ($SU(2)_L\times U(1)_Y$) are represented within square brackets. 
The additional fermionic Lagrangian in the present framework is therefore given as
\bea
{\cal L}_{Dark} &=& i\psi_{D_1}^\dagger {{\bar \sigma }^\mu }{D_\mu }{\psi_{D_1}} + i\psi_{D_2}^\dagger {{\bar \sigma }^\mu }{D_\mu }{\psi_{D_2}}+ \nonumber \\
&& i\psi_S^\dagger {{\bar \sigma }^\mu }{\partial_\mu }{\psi_S}-({m_{\psi}} \epsilon^{ab} {\psi_{D_1a}}{\psi_{D_2b}}+\frac{1}{2}c\phi\psi_S\psi_S+ h.c.)\,\, ,
\label{eq2}
\eea
where $D_\mu$ is the gauge covariant derivative in the Standard Model, 
$D_\mu =\partial_\mu - ig W_\mu^a \frac{\sigma_a}{2} - i g' Y B_\mu$.
From Eq.(\ref{eq2}) it can be easily observed that after $\phi$ gets a vev, the 
singlet fermion $\psi_S$ in the present model receives a Majorana mass, $m_{\psi_S}= c \langle \phi \rangle$. 

Apart from the interaction with the singlet scalar $\phi$, the dark sector doublet $\psi_{D_1}$ and singlet 
$\psi_S$ can also have Yukawa interactions with the Standard Model Higgs doublet, $H$. This Yukawa interaction term is 
given as
\begin{equation}
-{{\cal L}_{{\rm{Y}}}} = {\lambda}\psi_S\psi_{D_1}{H} + h.c.\,\, .
\label{eq3}
\end{equation}
Note that due to $Z_4$ charge assignment, $\psi_{D_2}$ does not have such Yukawa coupling in this present scenario.
Once $\phi$ gets a vev $v_{\phi}$ and the electroweak symmetry is broken (with $H$ acquires a vev $v/{\sqrt{2}}$, 
with $v=246$ GeV), 
Eq.(\ref{eq3}) generates a Dirac mass term for the additional neutral fermions. Hence including Eq.(\ref{eq2}) and 
Eq.(\ref{eq3}), the following mass matrix (involving the neutral fermions only) results 
\bea
\mathcal{M} &=& \left(\begin{array}{ccc} 
m_{\psi_S} & \frac{1}{\sqrt{2}}\lambda v &  0 \\
\frac{1}{\sqrt{2}}\lambda v & 0 & m_{\psi}\\
  0 & m_{\psi} & 0
\end{array}
\right).
\label{eq4}
\eea 
The matrix is constructed with the basis $ \mathcal{X}^T = (\psi_S,~\psi_1^0,~\psi_2^0)$. On the other hand, the charged components 
have a Dirac mass term, $m_{\psi} \psi_1^- \psi_2^+  + h.c.$. 

In general the mass matrix $\mathcal{M}$ could be complex. However for simplicity, we consider the parameters 
$m_{\psi_S}, \lambda$ and $m_{\psi}$ to be real. By diagonalizing this neutral fermion mass matrix, we obtain 
$V^T \mathcal{M} V = diag (m_{\chi_1}, m_{\chi_2}, m_{\chi_3})$, where the three physical states 
$\mathcal{P}^T = (\chi_1, \chi_2, \chi_3)$ are related to $\mathcal{X}$ by, 
\begin{equation}
\mathcal{X}_i = V_{ij} \mathcal{P}_j,
\label{eq:trans}
\end{equation}
where $V$ is diagonalizing matrix of $\mathcal{M}$. 
{Then the corresponding real mass eigenvalues obtained at the tree level are given as \cite{Enberg:2007rp,Adhikary:2013bma}}
\begin{align}
m_{\chi_1}&=-\frac{B}{3A}-\frac{2}{3A}\Big(\frac{R}{2}\Big)^{1/3} \cos\theta_m\, ,\label{eigenV1}\\
m_{\chi_2}&=-\frac{B}{3A}+ \frac{1}{3A}\Big(\frac{R}{2}\Big)^{1/3} (\cos\theta_m-\sqrt{3}\sin\theta_m)\, ,\label{eigenV2}\\
m_{\chi_3}&=-\frac{B}{3A}+ \frac{1}{3A}\Big(\frac{R}{2}\Big)^{1/3} (\cos\theta_m+\sqrt{3}\sin\theta_m)\, ,
\label{eigenV3}
\end{align}
where $A=1,~B=-m_{\psi_S},~C=-(m_{\psi}^2+\frac{\lambda^2v^2}{2}),
~D=m_{\psi}^2m_{\psi_S}$ (provided the discriminant ($\Delta$) of $\mathcal{M}$ is positive). Now $R$ and the angle $\theta_m$ can be  expressed as
\begin{align}
R=\sqrt{P^2+Q^2},~~~~~\tan3\theta_m=\frac{Q}{P}\,\, ,
\end{align}
where $P=2B^3-9ABC+27A^2D$ and $Q=3\sqrt{3\Delta}A$, $\Delta =18ABCD-4B^3D+B^2C^2-4AC^3-27A^2D^2$ is the
discriminant of the matrix $\mathcal{M}$. 
The lightest neutral fermion, protected by the unbroken $Z_2$,
can serve as a potential candidate for dark matter. 

At this stage, one can form usual four component spinors out of these physical fields \cite{Mahbubani:2005pt,DEramo:2007anh,Enberg:2007rp,Cohen:2011ec}. 
Below we define the Dirac fermion ($F^+$) and three neutral Majorana fermions ($F_{i=1,2,3}$) as,
\be
F^+ = \left ( \begin{array}{c} {\psi^+}_{\alpha} \\ (\psi^-)^{\dagger^{\dot{\alpha}}} \end{array} \right ),~~
F_{i} = \left ( \begin{array}{c} {\chi_i}_{\alpha} \\ (\chi_i)^{\dagger^{\dot{\alpha}}} \end{array} \right )\,\,,
\label{eq6}
\ee  
where $\psi_1^-$ and $\psi_2^+$ are identified with $\psi^-$ and $\psi^+$ respectively.  
In the above expressions of $F^+$ and $F_{i}$, $\alpha(\dot{\alpha})=1,2$ refers to 
upper (lower) two components of the Dirac spinor
that distinguishes the left handed Weyl spinor from the right handed Weyl spinor \cite{Martin:1997ns}.
Hence  $m_{F^+}=-m_{\psi}$ corresponds to the tree level Dirac 
mass for the charged fermion. 
Masses of the neutral fermions are then denoted as $m_{F_{i}} = m_{\chi_i}$. 
As we have discussed earlier, although the $Z_4$ symmetry is broken by $\langle\phi\rangle$,
a remnant $Z_2$ symmetry prevails in the dark sector which prevents 
dark sector fermions to have direct interaction with SM fermions. This can be understood later from the 
Lagrangian of Eq.(\ref{eq7}) which remains invariant if the dark sector fermions 
are odd under the remnant $Z_2$ symmetry.

Now we need to proceed for finding out various interaction terms involving these fields which 
will be crucial in evaluating DM relic density and finding direct detection cross-sections. However 
as in our model, there exists an extra singlet scalar, $\phi$ with non-zero vev, its mixing with SM 
Higgs doublet also requires to be included. For that purpose, we now discuss the scalar sector 
of our framework.

\subsection{Scalar sector}
\label{scalar}
As mentioned earlier, we introduce an additional real singlet scalar $\phi$ that carries $Z_4$ charge 
as given in Table~\ref{tab1}. The most general potential involving the SM Higgs doublet and 
the newly introduced scalar is given as  
\begin{align}
\mathcal{L_{\textrm{scalar}}}(H,\phi)=-\mu_H^2|H|^2+\lambda_H|H|^4-\frac{\mu_\phi^2}{2}\phi^2+\frac{\lambda_\phi}{4}\phi^4+\frac{\lambda_{\phi H}}{2}|H|^2 \phi^2.
\label{totP}
\end{align}
After electroweak symmetry is broken and $\phi$ gets vev, 
these scalar fields can be expressed as
\bea
H = \left( \begin{array}{c}
                          0  \\
        \frac{1}{\sqrt{2}}(v+H_0)  
                 \end{array}  \right) \, ,                     
~~~~~\phi=v_{\phi}+\phi_0 \,\, .
\label{fields}
\eea  
Minimization of the scalar potential leads to the following vevs of $\phi$  and $H$ given by 
\begin{align}
v_\phi^2=\frac{4\mu_\phi^2\lambda_H-2\mu_H^2\lambda_{\phi H}}{4\lambda_H\lambda_\phi-\lambda_{\phi H}^2}, \\
v^2=\frac{4\mu_H^2\lambda_\phi-2\mu_\phi^2\lambda_{\phi H}}{4\lambda_H\lambda_\phi-\lambda_{\phi H}^2}.
\end{align}
Therefore, after $\phi$ gets the vev and electroweak symmetry is broken, the mixing between the neutral component of 
$H$ and $\phi$ will take place (the mixing is parametrized by angle $\theta$) and new mass or physical eigenstates 
will be formed. The two physical eigenstates ($H_1$ and $H_2$)
can be obtained in terms of $H_0$ and $\phi_0$ as 
\begin{align}\label{eq:eigen}
 H_1=H_0 \cos\theta-\phi_0 \sin\theta,\nonumber\\
 H_2=H_0 \sin\theta+\phi_0 \cos\theta,
\end{align}
where $\theta$ is the scalar mixing angle defined by
\begin{align}\label{tanth}
\tan2\theta=\frac{\lambda_{\phi H} v v_\phi}{-\lambda_H  v^2+\lambda_{\phi } v_{\phi}^2}. 
\end{align}
{Similarly the mass eigenvalues of these physical scalars at 
tree level are found to be}
 \begin{align}
 m_{H_1}^2=\lambda_\phi v_\phi^2(1-\sec2\theta)+\lambda_Hv^2(1+\sec2\theta)\label{massE1},\\
 m_{H_2}^2=\lambda_\phi v_\phi^2(1+\sec2\theta)+\lambda_Hv^2(1-\sec2\theta)\label{massE2}.
 \end{align}
Using Eqs.(\ref{tanth}-\ref{massE2}), the couplings $\lambda_H$, $\lambda_\phi$ 
and $\lambda_{\phi H}$ can be expressed in terms 
of the masses of the physical eigenstates $H_1$ and $H_2$, the vevs ($v$, $v_\phi$) and the mixing angle 
$\theta$ as 
\begin{align}
\lambda_H =&\frac{m_{H_1}^2}{4 v^2}(1+\cos 2\theta)+\frac{m_{H_2}^2}{4 v^2}(1-\cos2\theta)\label{lambdaH},\\
\lambda_\phi=&\frac{m_{H_1}^2}{4v_\phi^{2}}(1-\cos 2\theta)+\frac{m_{H_2}^2}{4 v_{\phi}^ 2}(1+\cos2\theta)\label{lambdaphi},\\
\lambda_{\phi H}=&\sin2\theta\Big(\frac{m_{H_2}^2-m_{H_1}^2}{2 v v_\phi}\Big)\label{lambdaChiH}.
\end{align}
{Note that with $H_1$ as the SM Higgs, second term in Eq.~(\ref{lambdaH}) serves as the threshold correction to the SM Higgs quartic coupling. This would help $\lambda_H$ to maintain its positivity
at high scale.}  
Before proceeding for discussion 
of how this model  works in order to provide a successful DM scenario
and the status of electroweak vacuum stability, we first summarize relevant part of the interaction 
Lagrangian and the various vertices relevant for DM phenomenology and study of our model.

\subsection{Interactions in the model}

Substituting the singlet and doublet fermion fields of Eqs.(\ref{eq2}-\ref{eq3}) in terms of their 
mass eigenstates following Eq.(\ref{eq:trans}) and using the redefinition of fields given in 
Eq.(\ref{eq6}), gauge and Yukawa interaction terms can be obtained as 
{\small
\begin{eqnarray}
	\mathcal{L}_{int}&=& e A_\mu \bar{F^+} \gamma^\mu F^+
	+ \frac{g}{ 2 c_\mathrm{W}} (c_\mathrm{W}^2 - s_\mathrm{W}^2) Z_\mu \bar{F^+} \gamma^\mu F^+
	+\frac{g}{\sqrt{2}} \sum_i W_\mu^- (V_{3i}^* \bar{F}_{i} \gamma^\mu P_\mathrm{L} F^+ - V_{2i} \bar{F}_{i} \gamma^\mu P_\mathrm{R} F^+) \nonumber\\
	&&+ \frac{g}{\sqrt{2}} \sum_i W_\mu^+ (V_{3i} \bar{F^+} \gamma^\mu P_\mathrm{L} F_{i} - V_{2i}^* \bar{F^+} \gamma^\mu P_\mathrm{R} F_{i})
	-\frac{1}{2} \sum_{ij} \textrm{Re}X_{ij} Z_\mu \bar{F}_{i} \gamma^\mu \gamma^5 F_{j} 
	  +\frac{1}{2} \sum_{ij} \textrm{Im}X_{ij} Z_\mu \bar{F}_{i} \gamma^\mu F_{j} \nonumber\\
	&& -\frac{1}{2} \sum_{i} \bar{F_i} F_{i} [\textrm{Re}Y_{ii}\cos\theta-\textrm{Re}(cV_{1i}^2)\sin\theta]H_1
	-\frac{1}{2} \sum_{i\neq j}\bar{F_i} F_{j}[\textrm{Re}(Y_{ij}+Y_{ji})\cos\theta-\textrm{Re}(cV_{1i}V_{1j})\sin\theta]H_1
	 \nonumber
	\\&& -\frac{1}{2} \sum_{i} \bar{F_i} F_{i} [\textrm{Re}Y_{ii}\sin\theta+\textrm{Re}(cV_{1i}^2)\cos\theta]H_2
	-\frac{1}{2} \sum_{i\neq j}\bar{F_i} F_{j}[\textrm{Re}(Y_{ij}+Y_{ji})\sin\theta+\textrm{Re}(cV_{1i}V_{1j})\cos\theta]H_2 \nonumber \\
	&& +\frac{1}{2} \sum_{i} \bar{F_i} i \gamma^5 F_{i} [\textrm{Im}Y_{ii}\cos\theta-\textrm{Im}(cV_{1i}^2)\sin\theta]H_1 
	+\frac{1}{2} \sum_{i} \bar{F_i} i \gamma^5 F_{i} [\textrm{Im}Y_{ii}\sin\theta+\textrm{Im}(cV_{1i}^2)\cos\theta]H_2 \nonumber 
	\\&&+\frac{1}{2} \sum_{i\neq j} \bar{F_i} i \gamma^5 F_{j}[\textrm{Im}(Y_{ij}+Y_{ji})\cos\theta-\textrm{Im}(cV_{1i}V_{1j})\sin\theta]H_1 \nonumber
	\\&&+\frac{1}{2} \sum_{i\neq j} \bar{F_i} i \gamma^5 F_{j}[\textrm{Im}(Y_{ij}+Y_{ji})\sin\theta+\textrm{Im}(cV_{1i}V_{1j})\cos\theta]H_2\,\, .
\label{eq7}
\end{eqnarray}} 
Here the expressions of different couplings are given as
\begin{eqnarray}
	X_{ij} &=& \frac{g}{2 c_\mathrm{W}} (V_{2i}^* V_{2j} - V_{3i}^* V_{3j}),\quad \\
	Y_{ii} &=& \sqrt{2}~\lambda V_{1i} V_{2i},\quad
	Y_{ij}+Y_{ji} = \sqrt{2}~\lambda (V_{1i} V_{2j} + V_{1j} V_{2i}).
\label{eq8}
\end{eqnarray}
With the consideration that all the couplings involved in $\mathcal{M}$ are real, the elements of 
diagonalizing matrix $V$ in Eq.(\ref{eq:trans}) become real\cite{Adhikary:2013bma} and hence the interactions 
proportional to the imaginary parts in Eq.(\ref{eq7}) will disappear. Only real parts of 
$X_{ij},~Y_{ij}$ will survive. 
From Eq.(\ref{eq7}) we observe that the Largrangian remains invariant
if a $Z_2$ symmetry is imposed on the fermions. Therefore this residual $Z_2$ stabilises the lightest
fermion that serves as our dark matter candidate.

The various vertex factors involved in DM phenomenology, generated from the scalar Lagrangian, are
\begin{eqnarray}\label{Coups}
  H_1 f \bar{f}, H_2 f \bar{f}  & :&  \frac{m_f}{v} \cos\theta , \frac{m_f}{v} 
\sin\theta \nonumber \\
  H_1 Z Z, H_2 Z Z & :&  \frac{2 m_Z^2}{v} \cos\theta g^{\mu \nu}, \frac{2 
m_Z^2}{v} \sin\theta g^{\mu \nu}\nonumber \\  
  H_1 W^+W^-, H_2 W^+ W^- & :&  \frac{2 m_Z^2}{v} \cos\theta g^{\mu \nu}, \frac{2 m_Z^2}{v} \sin\theta g^{\mu \nu} \nonumber \\  
 H_1 H_1 H_1  &:& [6 v \lambda_H \cos^3\theta -3 v_\phi \lambda_{\phi H} \cos^2\theta \sin\theta +3 v \lambda_{\phi H} \cos\theta \sin^2\theta-6v_\phi \lambda_\phi \sin^3\theta]\nonumber \\
 H_2 H_2 H_2  &:& [6 v \lambda_H \sin^3\theta +3 v_\phi \lambda_{\phi H} \cos\theta \sin^2\theta +3 v \lambda_{\phi H} \cos^2\theta \sin\theta+6v_\phi \lambda_\phi \cos^3\theta]\nonumber \\
 H_1 H_1 H_2  &:& [2 v (3 \lambda_H - \lambda_{\phi H})\cos^2\theta \sin\theta +v \lambda_{\phi H} \sin^3\theta + v_\phi\nonumber (6\lambda_{\phi}-2\lambda_{\phi H})\cos\theta \sin^2\theta 
 \\ &&+v_\phi \lambda_{\phi H}\cos^3\theta ]\nonumber \\
 H_1 H_2 H_2  &:& [2 v (3 \lambda_H - \lambda_{\phi H})\cos\theta \sin^2\theta +v \lambda_{\phi H} \cos^3\theta - v_\phi (6\lambda_{\phi}-2\lambda_{\phi H})\cos^2\theta \sin\theta\nonumber
 \\&& -v_\phi \lambda_{\phi H}\sin^3\theta ],\nonumber \\
\label{couplings}
 \end{eqnarray}
 where {$m_f$ represents mass of SM fermion(s) ($f$) and $m_Z$ corresponds to the mass of the $Z$ boson (at tree level)}.



\section{Constraints}
\label{constraints}
In this section we illustrate important theoretical and experimental bounds that can constrain the parameter space of the proposed model. {Note that among $H_1$ and $H_2$, one of them would be the Higgs discovered at LHC (say the SM Higgs). The other Higgs can be heavier or lighter than the SM Higgs. In this analysis, we consider the lightest 
scalar state $H_1$ as the Higgs with mass $m_{H_1} =125.09$ GeV \cite{Patrignani:2016xqp}. We argue at the end of this section why such a choice is phenomenologically favoured from DM and vacuum stability issues
with respect to the case with additional Higgs being lighter than the SM one}. {Now from the discussion of 
Secs. \ref{fermion} and \ref{scalar}, it turns out that there are
six independent parameters in the set up: three ($m_{H_2}$, $\sin\theta$ and $v_\phi$) from the scalar sector and other three
 ($\lambda, m_\psi, c$) from the fermionic sector.}
These parameters can be constrained using the limits from perturbativity, perturbative unitarity,
electroweak precision data and the singlet induced NLO correction to the W boson mass \cite{Robens:2015gla,Chalons:2016lyk,Robens:2016xkb}. { In addition,
 constraints from DM experiments, LHC and LEP will also be applicable.  We discuss these constraints below.} 
 
\subsection*{[A] Theoretical Constraints:}
\begin{itemize}
\item The scalar potential should be bounded from below in any field direction.
 This poses some constraints\cite{Kannike:2012pe,Chakrabortty:2013mha} on the scalar couplings of the model which we 
 will discuss in Sec.~\ref{vs} in detail. The conditions must be satisfied at any energy scales till $M_P$ in order to ensure the stability of the entire scalar potential in any field direction. 
\item One should also consider the the perturbative unitarity bound associated with
the S matrix corresponding to scattering
processes involving all two-particle initial and final
states. In the specific model
under study, there are five neutral
 ($W^+W^-,~ZZ,~H_0H_0,~H_0\phi_0,~\phi_0\phi_0$) and three singly
charged ($W^+ H_0,~W^+\phi_0,~W^+Z$) combinations of two-particle initial 
{and} final
states\cite{Horejsi:2005da,Bhattacharyya:2015nca,Kang:2013zba}. The perturbative unitarity limit can be derived by implementing the
bound on the scattering amplitude $\mathcal{M}$\cite{Horejsi:2005da,Bhattacharyya:2015nca,Kang:2013zba}
\begin{align}
 \mathcal{M}<8\pi.
\end{align}
The unitarity constraints are obtained as \cite{Horejsi:2005da,Bhattacharyya:2015nca,Kang:2013zba}
\begin{align}
 \lambda_H<4\pi,~~~\lambda_{\phi H}<8\pi,~~\textrm{and}~~\frac{1}{4}\{12\lambda_H+\lambda_\phi\pm\sqrt{16\lambda_{\phi H}^2+(\lambda_\phi-12\lambda_H)^2}\}<8\pi.
\end{align}
\item {In addition, all relevant couplings in the framework should maintain the perturbativity limit. 
Perturbative conditions of relevant couplings in our set up appear as\cite{Chakrabarty:2014aya} \footnote{With a Lagrangian term like $\lambda\phi_i\phi_j\phi_k\phi_l$, the perturbative expansion parameter for a $2\rightarrow 2$ process involving different scalars $\phi_{i,j,k,l}$
turns out to be $\lambda$. Hence the limit is $\lambda<4\pi$ \cite{Chakrabarty:2014aya} . Similarly with a term $yS f_if_j$ involving scalar $S$ and fermions $f(i\neq j)$,
the corresponding expansion parameter is restricted by $y^2<4\pi$ \cite{Chakrabarty:2014aya} .  Considering the associated symmetry factors (due to the presence of identical fields), we arrive at the limits mentioned at Eq.(\ref{PerL}).}}
\begin{align}
\lambda_H<\frac{2}{3}\pi,~\lambda_\phi<\frac{2}{3}\pi,~\lambda_{\phi H}<4\pi, ~\lambda<\sqrt{4\pi},~~\textrm{and}~c<\sqrt{4\pi}.\label{PerL}
\end{align}
We will ensure the perturbativity of
the couplings present in the model till $M_P$ energy scale by employing the renormalization group equations.
\end{itemize}

\subsection*{[B] Experimental Constraints:}
\begin{itemize}

\item In the present singlet doublet dark matter model, dark matter candidate $\chi_1$
has coupling with the Standard Model Higgs $H_1$ and neutral gauge boson $Z$.
Therefore, if kinematically allowed, the gauge boson and Higgs can decay into pair of dark matter particles.
Hence we should take into account the bound on invisible decay width of Higgs and $Z$ boson from
LHC and LEP. The corresponding {tree level} decay widths of Higgs 
boson $H_1$ and $Z$
into DM is given as
\bea
\Gamma_{H_1}^{inv}=\frac{\lambda_{H_1 {\chi}_{1} \chi_{1}}^2}{16 \pi}m_{H_1}
\left(1-\frac{4m_{\chi_{1}}^2}{m_{H_1}^2}\right)^{3/2}\, \nonumber \\
\Gamma_{Z}^{inv}=\frac{\lambda_{Z \chi_{1} \chi_{1}}^2}{24 \pi}m_{Z}
\left(1-\frac{4m_{\chi_{1}}^2}{m_{Z}^2}\right)^{3/2}\, ,
\label{inv}
\eea
where the couplings, $\lambda_{H_1 \chi_{1} \chi_{1}}$ and $\lambda_{Z \chi_{1} \chi_{1}}$
can be obtained from Eq.(\ref{eq7}). The bound on $Z$ invisible decay width from LEP
is $\Gamma_Z^{inv}\le2$ MeV at 95\% C.L. \cite{ALEPH:2005ab} while LHC provides
bound on Higgs invisible decay and invisible decay branching fraction
$\Gamma_{H_1}^{inv}/\Gamma_{H_1}$ is 23\% \cite{CMS:2016rfr}.

\item The mass of the SM gauge boson W gets correction from the scalar induced one loop diagram\cite{Lopez-Val:2014jva}. 
{This poses stronger limit on the scalar mixing angle $\sin\theta\lesssim (0.3-0.2)$ for $300~{\rm GeV}<m_{H_2}<800$ GeV\cite{Robens:2016xkb}.}

\item Moreover, the Higgs production cross-section
also gets modified in the present model due to mixing with the real
scalar singlet. As a result, Higgs production cross-section at
LHC is scaled by a factor $\cos^2\theta$ and the corresponding
Higgs signal strength is given as 
$R=\frac{\sigma_{H_1}}{\sigma_{SM}}\frac{Br(H_1\rightarrow XX)}{Br_{SM}}$ \cite{Strassler:2006ri},
where $\sigma_{SM}$ is the SM Higgs production cross-section and $Br_{SM}$
is the measure of SM Higgs branching ratio to final state particles $X$.
The simplified expression for the signal strength is 
given as \cite{Strassler:2006ri,Robens:2016xkb,Khachatryan:2015cwa,Aad:2015kna,Chatrchyan:2013mxa,CMS1,CMS2,ATLAS1}
\be
R=\cos^4\theta\frac{\Gamma_1}{\Gamma_{H_1}^{Tot}}\,\, ,
\ee
where $\Gamma_1$ is the decay width of $H_1$ in SM. In absence of any invisible decay (when $m_{\chi_1}>m_{H_1}/2$), the 
signal strength is simply given {as} $R=\cos^2\theta$. Since $H_1 $
is the SM like Higgs with mass $ 125.09$ GeV, $R\simeq1$. Hence, this
restricts the mixing between the scalars. The ATLAS\cite{ALEPH:2005ab} and CMS\cite{CMS:2016rfr} combined result provides
\begin{align}
R=1.09^{+0.11}_{-0.10}.
\end{align}
This can be translated into an upper bound on $\sin\theta \lesssim 0.36$ at $3\sigma$.

\hspace{3mm}     Similarly, one can also obtain signal strength of the other 
scalar involved in the model expressed as 
$R^{'}=\sin^4\theta\frac{\Gamma_2}{\Gamma_{H_2}^{Tot}}$,
where $\Gamma_2$ being the decay with of $H_2$ with mass $m_{H_2}$ in SM and
$\Gamma_{H_2}^{Tot}$ is the total decay width of the scalar $H_2$ given as
$\Gamma_{H_2}^{Tot}=\sin^2\theta~\Gamma_2+\Gamma_{H_2}^{inv}+
{\Gamma_{H_2\rightarrow H_1H_1}}$. {The additional 
term $\Gamma_{H_2\rightarrow H_1H_1}$ appears when $m_{H_2}\geq2m_{H_1}$ and is 
expressed as 
$\Gamma_{H_2\rightarrow 
H_1H_1}=\frac{\lambda_{H_1H_1H_2}^2}{32\pi 
m_{H_2}}\sqrt{1-\frac{4m_{H_1}^2}{m_{H_2}^2 }} $, where $\lambda_{H_1H_1H_2}$ 
can be obtained from Eq.(~\ref{couplings}).} 
However due to
small mixing with the SM Higgs $H_1$, $R^{'}$ is very small to provide any
significant signal to be detected at LHC\cite{Robens:2016xkb}.

{\item In addition, we  include the LEP bound on the charged fermions involved in the singlet doublet model. The present limit from LEP excludes a singly charged fermion having mass below 100 
GeV \cite{Abdallah:2003xe}. Therefore we consider $m_{\psi}\gtrsim$ 100 GeV. The LEP bound on the heavy Higgs state  (having mass above 250 GeV)  turns out to be weaker  compared to the limit obtained from $W$ boson mass correction \cite{Robens:2016xkb}.} 

\item The presence of fermions in the dark sector and the additional scalar $\phi$ will affect
the oblique parameters\cite{Peskin:1991sw} $S$, $T$ and $U$ through changes in gauge boson propagators.
However only $T$ parameter could have a relevant contributions from the newly introduced fields. Contributions
to the $T$ parameter by the additional scalar field $\phi$ can be found in \cite{Barger:2007im}. 
However in the small mixing case, this turns out to be negligible \cite{Ghosh:2015apa} and
can be safely ignored\cite{Enberg:2007rp}.
When we consider fermions, the corresponding $T$ parameter
in our model is obtained as \cite{Barbieri:2006bg,DEramo:2007anh} 
\begin{align}
 \Delta T=\sum_{i=1}^{3}\Big[\frac{1}{2}(V_{3i}-V_{2i})^2\mathcal{A}(m_\psi,m_i)+\frac{1}{2}(V_{3i}+V_{2i})^2\mathcal{A}(m_\psi,-m_i)\Big]\nonumber\\
-\sum_{i,j=1}^{3}\frac{1}{4}(V_{2i}V_{2j}-V_{3i}V_{3j})^2\mathcal{A}(m_i,-m_j),
\end{align}
where $\mathcal{A}(m_i,m_j)=\frac{1}{32 \alpha_{\textrm{em}}\pi v^2}\Big[(m_i-m_j)^2 \textrm{ln}\frac{\Lambda^4}{m_i^2m_j^2}-2 m_im_j+\frac{2m_im_j(m_i^2+m_j^2
)-m_i^4-m_j^4}{m_i^2-m_j^2}\textrm{ln}\frac{m_i^2}{m_j^2}\Big]$ and $\Lambda$ is the cutoff of the 
loop integral which vanishes during the numerical estimation.

\item {Furthermore, we also use 
the measured value of DM relic abundance by Planck experiment \cite{Ade:2015xua}  and apply limits on DM direct detection cross-sections from LUX \cite{Akerib:2016vxi}, XENON-1T  \cite{Aprile:2017iyp}, Panda 2018 \cite{Cui:2017nnn} and XENON-nT \cite{Aprile:2015uzo} experiments to constrain the parameter space of the model. Detailed discussions on direct 
searches of dark matter have been presented later in Sec.~\ref{dm}.} 
\end{itemize}

{In the above discussion, we infer that the scalar mixing angle $\sin\theta$ is restricted by $\sin\theta\lesssim0.3$, provided the mass of additional Higgs ($m_{H_2}$) is around 300 GeV. For further heavier $m_{H_2}$, $\sin\theta$ is even more restricted, {\it e.g.} $\sin\theta\lesssim 0.2$ for $m_{H_2}$ around 800 GeV. On the other hand, if we consider $H_1$ to be lighter than the Higgs discovered at LHC, we need to identify $H_2$ as the SM Higgs as per Eqs.(\ref{eq:eigen},\ref{massE1}-\ref{massE2}) (where $\sin\theta\rightarrow 1$ is the decoupling limit). In this case, the limit turns out to be $\sin\theta\gtrsim 0.87$ for $m_{H_1}\lesssim 100$ GeV \cite{Robens:2016xkb}. Note that this case is not interesting from vacuum stability point of view in this work for the following reason. From Eq.~(\ref{lambdaH}), we find the first term in right hand side serves as the threshold correction to the SM Higgs quartic coupling (contrary to the case with $H_1$ as the SM 
Higgs and $H_2$ as the heavier one, where the threshold correction is provided by the second term). However with $m_{H_1}<m_{H_2}\equiv$ SM Higgs and $\sin\theta\gtrsim 0.87$, the contribution of the first term is much less compared to the second term. Hence in this case, the SM Higgs quartic coupling $\lambda_{H}$ cannot be enhanced significantly such that its positivity till very high scale can be ensured\footnote{With $m_{H_2}>m_{H_1}\equiv$ SM Higgs and $\sin\theta\sim 0.1-0.3$, the second term can definitely contribute to a large extent toward the positivity of $\lambda_{H}$.}. Therefore we mainly focus on the case with $m_{H_2}>m_{H_1}(\equiv$ SM Higgs) for the rest of our analysis.}

\section{Dark matter phenomenology}
\label{dm}

In the present model, apart from the SM particles we have three neutral Majorana fermions,
one charged Dirac fermion and one additional Higgs (other than the SM one). Out of these, 
the lightest neutral Majorana ($\chi_1$) plays the role of dark matter. Being odd under residual $Z_2$, 
stability of the DM is ensured. As observed through Eq.(\ref{eq4}), masses of these neutral 
Majorana fermions depend effectively on three parameters  $m_{\psi_S},~\lambda$ and 
$m_{\psi}$. However in our present scenario, $m_{\psi_S}$ actually involves two parameters; $c$ 
and $v_{\phi}$, the individual roles of which are present in DM annihilation and 
vacuum stability. 
For the case when coupling $\lambda$ is small ($\lambda < 1$), with $m_{\psi}>m_{\psi_S}$ (with 
$\lambda v/{\sqrt{2}} < m_{\psi}$), 
our DM candidate remains singlet dominated and for $m_{\psi}<m_{\psi_S}$, this becomes 
doublet like \cite{Xiang:2017yfs}.
In the present work we will investigate the characteristics of the dark matter candidate irrespective 
of its singlet or doublet like nature. 
\subsection{Dark Matter relic Density} 
\begin{figure}[H]
 \centering
 \includegraphics[height=8.2cm, width=11 cm,angle=0]{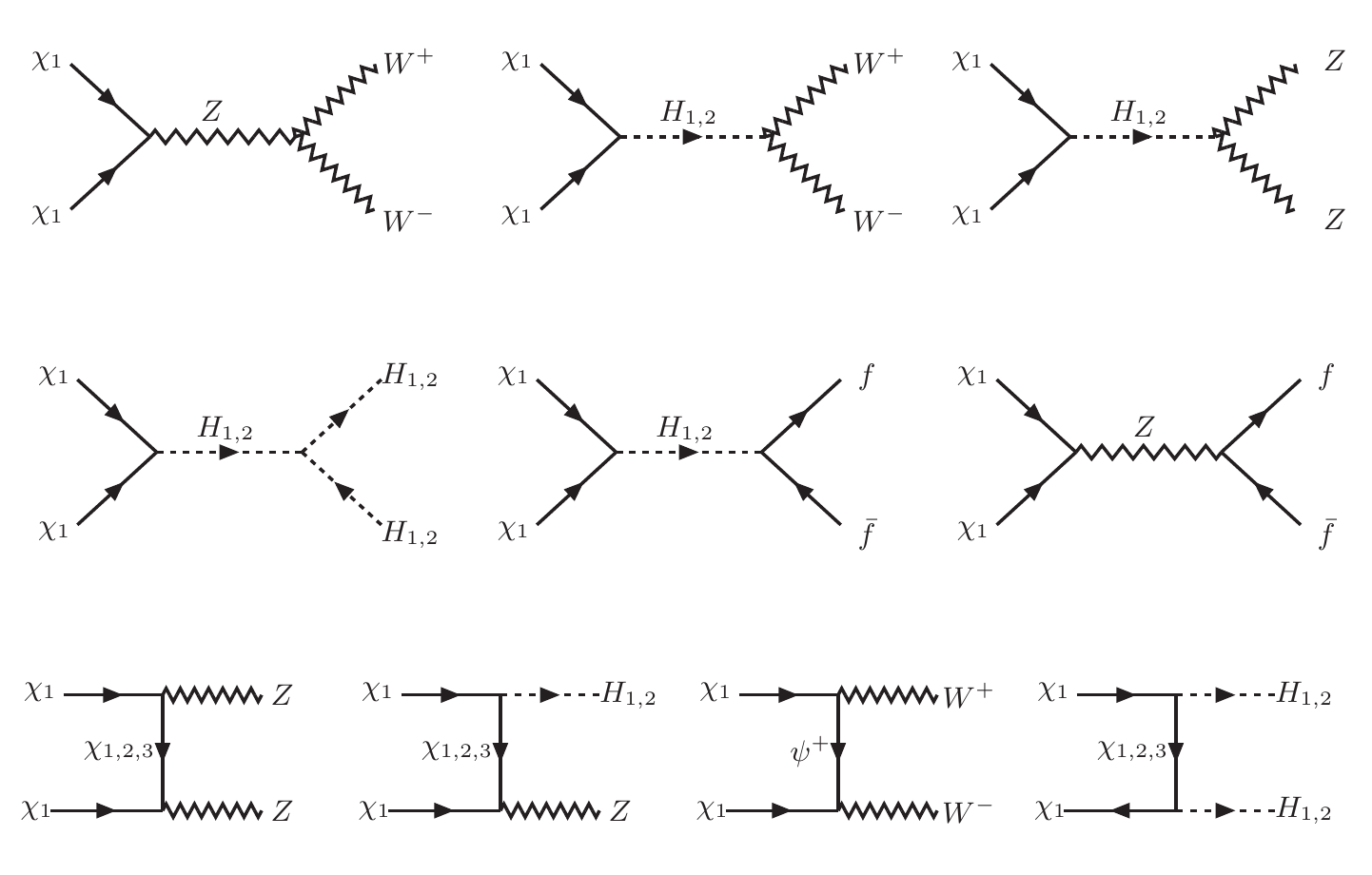}\\
 \caption{The dominant annihilation channels of DM to SM fields and heavy Higgs in the final states.}
 \label{feyn1}
 \end{figure}

\begin{figure}[H]
 \centering
 \includegraphics[height=9.5 cm, width=11 cm,angle=0]{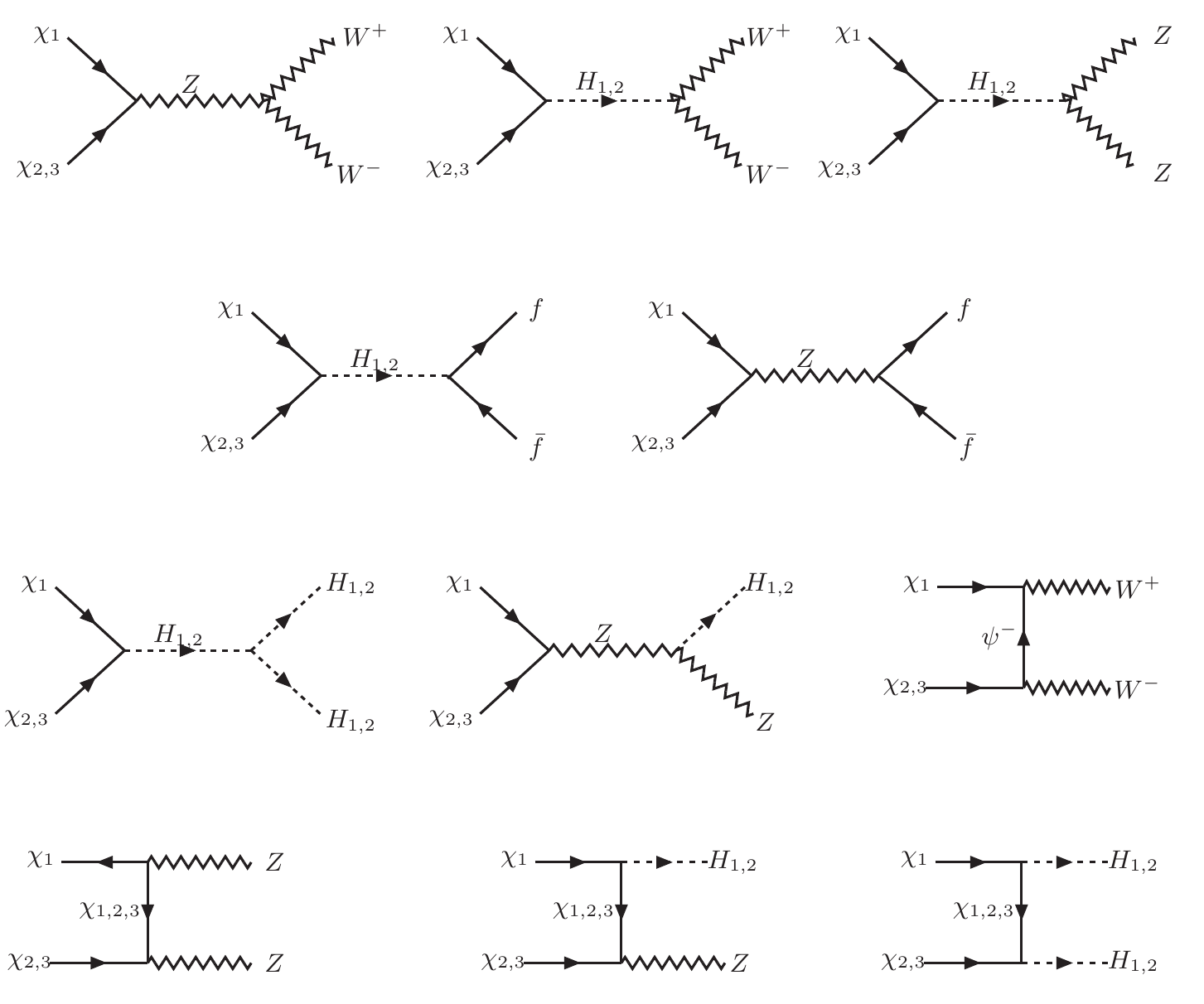}\\
 \caption{The dominant co-annihilation channels of DM (${\chi}_1$) with neutral fermions ${\chi}_{2,3}$.}
 \label{feyn2}
\end{figure}

\begin{figure}[H]
 \centering
 \includegraphics[height=6.0 cm, width=11 cm,angle=0]{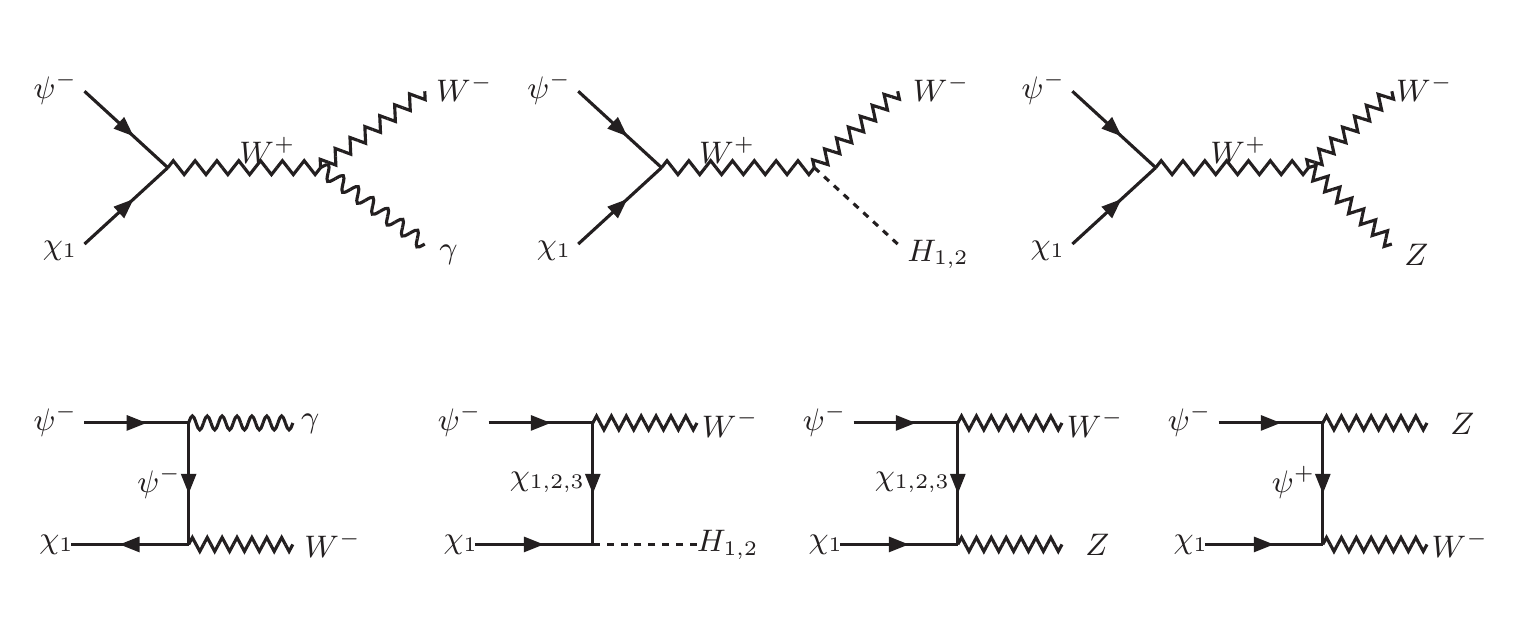}\\
 \caption{The dominant co-annihilation channels of DM (${\chi}_1$) with charged fermion $\psi^-$.}
 \label{feyn3}
\end{figure}

\begin{figure}[H]
 \centering
 \includegraphics[height=7.5 cm, width=11cm,angle=0]{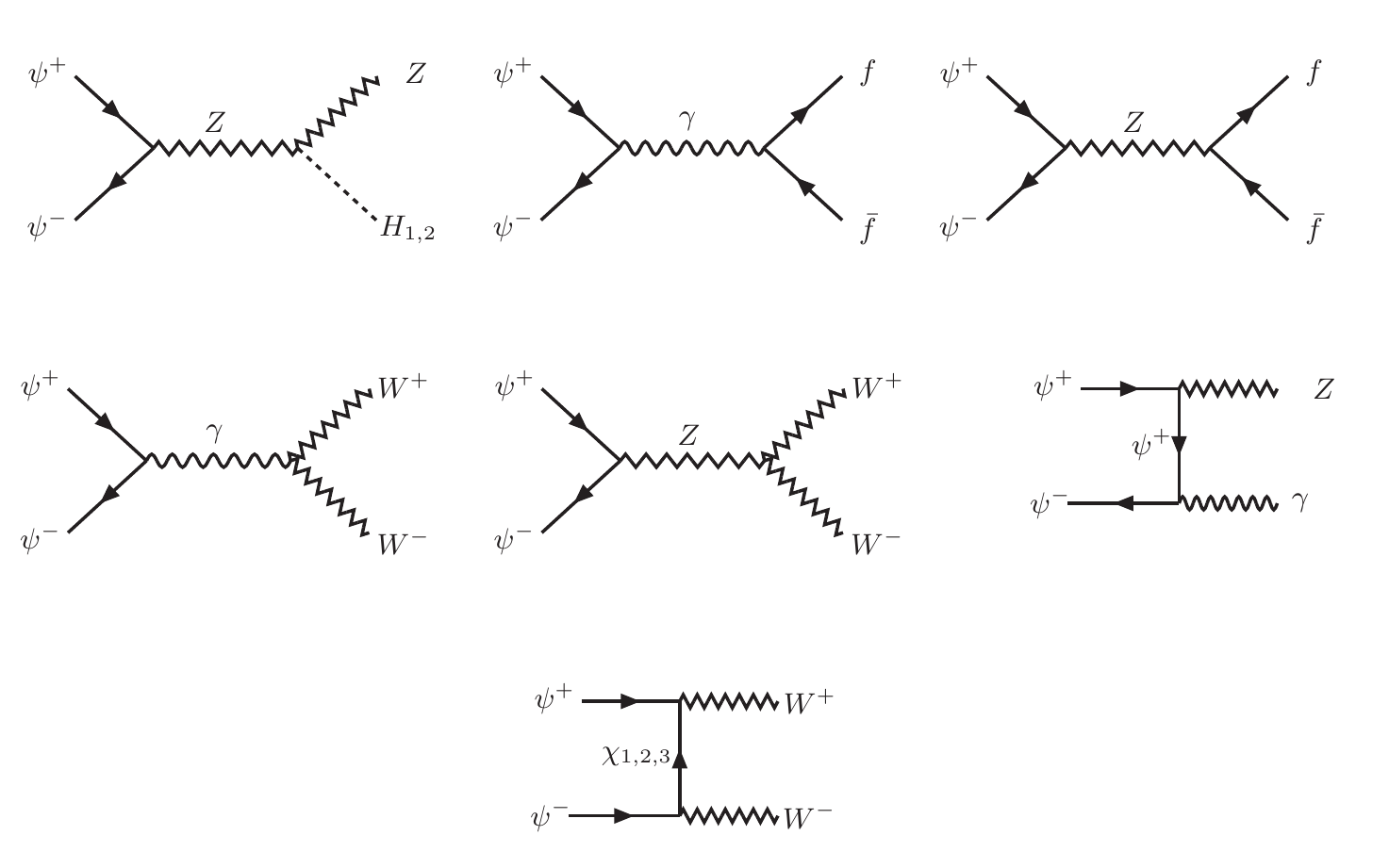}\\
 \caption{The dominant co-annihilation channels of the charged fermion pair $\psi^+$ and $\psi^-$.}
 \label{feyn4}
\end{figure}

 Dark matter relic density
is obtained by solving the Boltzmann equation. The expression for dark matter relic
density is given as \cite{McDonald:1993ex,Cynolter:2008ea}
\be
\Omega_{\chi_1}h^2=\frac{2.17\times10^8~{\rm GeV}^{-1}}{g_{\star}^{1/2}M_{P}}\frac{1}{J(x_f)}\,\, ,
\label{relic}
\ee
where $M_{P}$ denotes the reduced Planck mass ($2.435\times 10^{18}$ GeV) and the factor $J(x_f)$ is expressed as
\be
J(x_f)=\int_{x_f}^{\infty}\frac{\langle \sigma |v| \rangle}{x^2}dx\, \, ,
\label{sigmav}
\ee 
where $x_f=m_{\chi_1}/T_f$, with $T_f$ denoting freeze out temperature and $g_{\star}$ 
is the total number of degrees of freedom of particles.  
In the above expression, $\langle \sigma |v| \rangle$ is the measure of thermally averaged annihilation cross-section of dark matter $\chi_1$ into different SM final state
particles. It is to be noted that annihilation of dark matter in the present model also
includes co-annihilation channels due to the presence of other dark sector particles. Different 
Feynmann diagrams for dark matter annihilations and co-annihilations are shown in 
Fig.~\ref{feyn1} and Figs.~\ref{feyn2},\ref{feyn3},\ref{feyn4} respectively.
\noindent 

The thermally averaged dark matter annihilation cross-section $\langle \sigma |v| \rangle$ is expressed as
{\small
\begin{equation}
\begin{split}
\langle \sigma |v| \rangle= & \frac{g_1^{'2}}{g_{eff}^2} \sigma (\chi_1 \chi_1)
 +2 \frac{g'_1 g'_2}{g_{eff}^2} \sigma (\chi_1 {\chi}_2) (1+\Delta_{21})^{3/2} exp(-x\Delta_{21})
 +2 \frac{g'_1 g'_3}{g_{eff}^2} \sigma (\chi_1 {\chi}_3) (1+\Delta_{31})^{3/2} exp(-x\Delta_{31})\\
 &+2 \frac{g'_2 g'_3}{g_{eff}^2} \sigma (\chi_2 {\chi}_3) (1+\Delta_{21})^{3/2}(1+\Delta_{31})^{3/2} exp(-x(\Delta_{21}+\Delta_{31}))\\
 &+2 \frac{g'_1 g'_+}{g_{eff}^2} \sigma (\chi_1 \psi^+) (1+\Delta_{+1})^{3/2} exp(-x\Delta_{+1})
 +\frac{g_+^{'2}}{g_{eff}^2} \sigma (\psi^+ \psi^-) (1+\Delta_{+1})^{3} exp(-2x\Delta_{+1})\\
 &+2 \frac{g'_2 g'_+}{g_{eff}^2} \sigma (\chi_2 \psi^+) (1+\Delta_{+1})^{3/2}(1+\Delta_{21})^{3/2} exp(-x(\Delta_{+1}+\Delta_{21}))\\
 &+2 \frac{g'_3 g'_+}{g_{eff}^2} \sigma (\chi_3 \psi^+) (1+\Delta_{+1})^{3/2}(1+\Delta_{31})^{3/2} exp(-x(\Delta_{+1}+\Delta_{31})\\
 &+ \frac{g_2^{'2}}{g_{eff}^2} \sigma (\chi_2 {\chi}_2) (1+\Delta_{21})^{3} exp(-2x\Delta_{21})
 +\frac{g_3^{'2}}{g_{eff}^2} \sigma (\chi_3 {\chi}_3) (1+\Delta_{31})^{3} exp(-2x\Delta_{31})\,\, ,
\end{split}
\end{equation}}
where $\Delta_{i1}=\frac{{m_{\chi_i}}-m_{\chi_1}}{m_{\chi_1}}$ and $\Delta_{+1}=\frac{m_{\psi}-m_{\chi_1}}{m_{\chi_1}}$
are the corresponding mass splitting ratios. Therefore it can be easily concluded that for smaller values of mass splitting
co-annihilation effects will enhance the final dark matter annihilation cross-section significantly. The effective
degrees of freedom $g_{eff}$ 

\noindent 
is denoted as
\be
g_{eff} = g'_1+ g'_2 (1+\Delta_{21})^{3/2} exp(-x\Delta_{21}) + g'_3 (1+\Delta_{31})^{3/2} exp(-x\Delta_{31})
+ g'_+ (1+\Delta_{+1})^{3/2} exp(-x\Delta_{+1})\,.
\label{dof}
\ee

In the above expression $g'_i,i=1-3$ are spin degrees of freedom of particles.
Using Eqs.(\ref{relic}-\ref{dof}), relic density of the dark matter ${\chi}_1$ can be obtained for 
the model parameters. The relic density of the dark matter candidate must satisfy the bounds from
Planck \cite{Ade:2015xua} with $ 1\sigma$ uncertainty is given as 
\be
0.1175\le\Omega_{DM}h^2\le0.1219\,\, .
\label{planck}
\ee

\subsection{Direct searches for dark matter}

 \begin{figure}[H]
 \centering
 \includegraphics[height=4.55 cm, width=4 cm,angle=0]{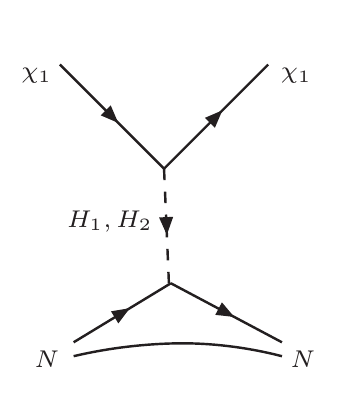}~~~~~~~~~~~~~~~~
 \includegraphics[height=4.4 cm, width=4 cm,angle=0]{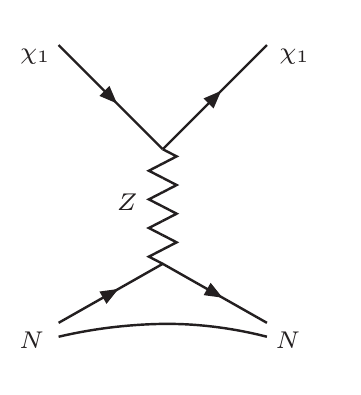}\\
 \caption{Schematic diagrams for dark matter direct detection processes: left panel: spin independent and right panel:
spin dependent processes ($N$ is the nucleon).}
 \label{DMDD}
 \end{figure}
Direct detection of dark matter is based on the scattering of the incoming dark
matter particle with detector nucleus.
In the present scenario,
the dark matter candidate $\chi_1$ can have both spin independent (SI)
and spin dependent (SD) scatterings with the detector. In view of Eq.(\ref{eq7}), spin independent
interactions are mediated by scalars $H_1$ and $H_2$ while spin dependent scattering
is mediated via neutral gauge boson $Z$  as shown in Fig.~\ref{DMDD}. 

The expression for spin independent direct detection cross-section
in the present singlet doublet model is given as \cite{Gabrielli:2013hma}
\bea
\sigma_{SI}\simeq\frac{m_r^2}{8\pi}
\left(\frac{\lambda_{H_1 \chi_1 \chi_1}\cos{\theta}}{m_{H_1}^2}-
\frac{\lambda_{H_2 \chi_1 {\chi}_1}\sin{\theta}}{m_{H_2}^2}\right)^2
\lambda_p^2
\label{dd}
\eea
where $\lambda_{H_i \chi_1 \chi_1},~i=1,2$ denotes the coupling
of dark matter ${\chi}_1$ with the scalar $H_1$ and $H_2$ as given in the
Eq.(\ref{eq7}). In the above expression of direct detection cross-section, $m_r$
is the reduced mass for the dark matter-nucleon scattering,
$m_r=\frac{m_{\chi_1} m_p}{m_{\chi_1}+m_p}$, $m_p$ being the proton
mass. 
The scattering factor $\lambda_p$ is expressed as \cite{LopezHonorez:2012kv}
\be
\lambda_p=\frac{m_p}{v}\left[\sum_q f_q+ \frac{2}{9}\left(1-\sum_q
f_q\right)\right] \simeq1.3\times10^{-3} \,\, ,
\label{lp}
\ee
where $f_q$ is the atomic form factor\cite{Alarcon:2011zs,Alarcon:2012nr}.

As we have mentioned earlier, following the interaction Lagrangian described
in Eq.(\ref{eq7}), we have an axial vector interaction of the neutral
Majorana fermions with the SM gauge boson $Z$. This will infer spin dependent
dark matter nucleon scattering with the detector nuclei. The expression
for the spin dependent cross-section is given as \cite{Agrawal:2010fh}
\begin{align}
   \sigma_{SD}
  =
  \frac{16 m_r^2}{\pi}
  \left[\sum_{q=u,d,s}{d_q\lambda_q}\right]^2
  J_N(J_N+1).
  \label{sd}
\end{align}
where $d_q\sim \frac{g^2}{2c_W m_Z^2}ReX_{11}$  (following Eq.(\ref{eq7})) and $\lambda_q$ 
depends on the nucleus considering ${\chi}_1$ as the dark matter candidate. 

\subsection{Results}
\label{res}

In this section we present the dark matter phenomenology involving different model parameters and constrain the parameter space with
theoretical and experimentally observed bounds discussed in Sec.~\ref{dm}.
As mentioned earlier, the dark matter candidate is a thermal WIMP (Weakly Interacting Massive Particle) 
in nature. The dark matter
phenomenology is controlled by the following parameters \footnote{Note that although $c$ and $v_{\phi}$ together forms $m_{\psi_S}$ appearing in neutral fermion mass eigenvalues (see Eq.~(\ref{eigenV1}-\ref{eigenV3})), the parameter $c$ alone ({\it i.e.} without $v_{\phi}$) is involved in DM-annihilation processes (see Eq.~(\ref{eq7})). Hence we treat both $c$ and $v_{\phi}$ as independent parameters.},
\be
\{c,~v_{\phi},~\lambda,~\sin\theta,~m_{\psi}\,, ~m_{H_2}\}. \nonumber
\ee
\noindent{We have used LanHEP (version 3.2) \cite{Semenov:2014rea} to 
extract the model files and use MicrOmegas (version 3.5.5)
\cite{Belanger:2013oya} to perform the numerical analysis.} The model in general 
consists of three neutral fermions ${\chi}_i,~i=1-3$ and one charged 
fermion $\psi^+$ which take part in this analysis. 
The lightest fermion ${\chi}_1$ is the dark
matter candidate that annihilates into SM particles and freeze out to provide the required
dark matter relic density. The heavier neutral particles in the dark sector ${\chi}_{2,3}$
and the charged particle $\psi$ annihilates into the lightest particle ${\chi}_1$. Also ${\chi}_{2,3}$
co-annihilation contributes to the dark matter relic abundance (when the mass differences are small). 
Different possible annihilation and co-annihilation channels of the dark matter
particle is shown in Figs.~\ref{feyn1}, \ref{feyn2}, \ref{feyn3}, \ref{feyn4} . 

We have kept the mass of the heavier Higgs $m_{H_2}$ below 1 TeV from the viewpoint of future experimental search at LHC. {In particular,} unless otherwise stated, for discussion purpose we have kept the heavy Higgs at 300 GeV.
{Also note that in this regime, $\sin\theta$ is bounded by $\sin\theta\lesssim 0.3$\cite{Robens:2016xkb}, so we could exploit maximum amount of variation for $\sin\theta$ as otherwise with heavier $H_2$ $\sin\theta$ will be more restrictive.}
In the small $\sin\theta$ approximation, $\lambda_\phi$ almost coincides with the second term in Eq.(\ref{lambdaphi}). 
 Now it is quite natural
to keep the magnitude of a coupling below unity to maintain the perturbativity at all energy
scales (including its running). Hence with the demand $\lambda_\phi < 1$, from Eq.(\ref{lambdaphi}) one finds 
$v_\phi >\sqrt{3}m_{H_2}$. 

\subsubsection{Study of importance of individual parameters}

\begin{figure}[H]
 \centering
 \includegraphics[height=7 cm, width=7 cm,angle=0]{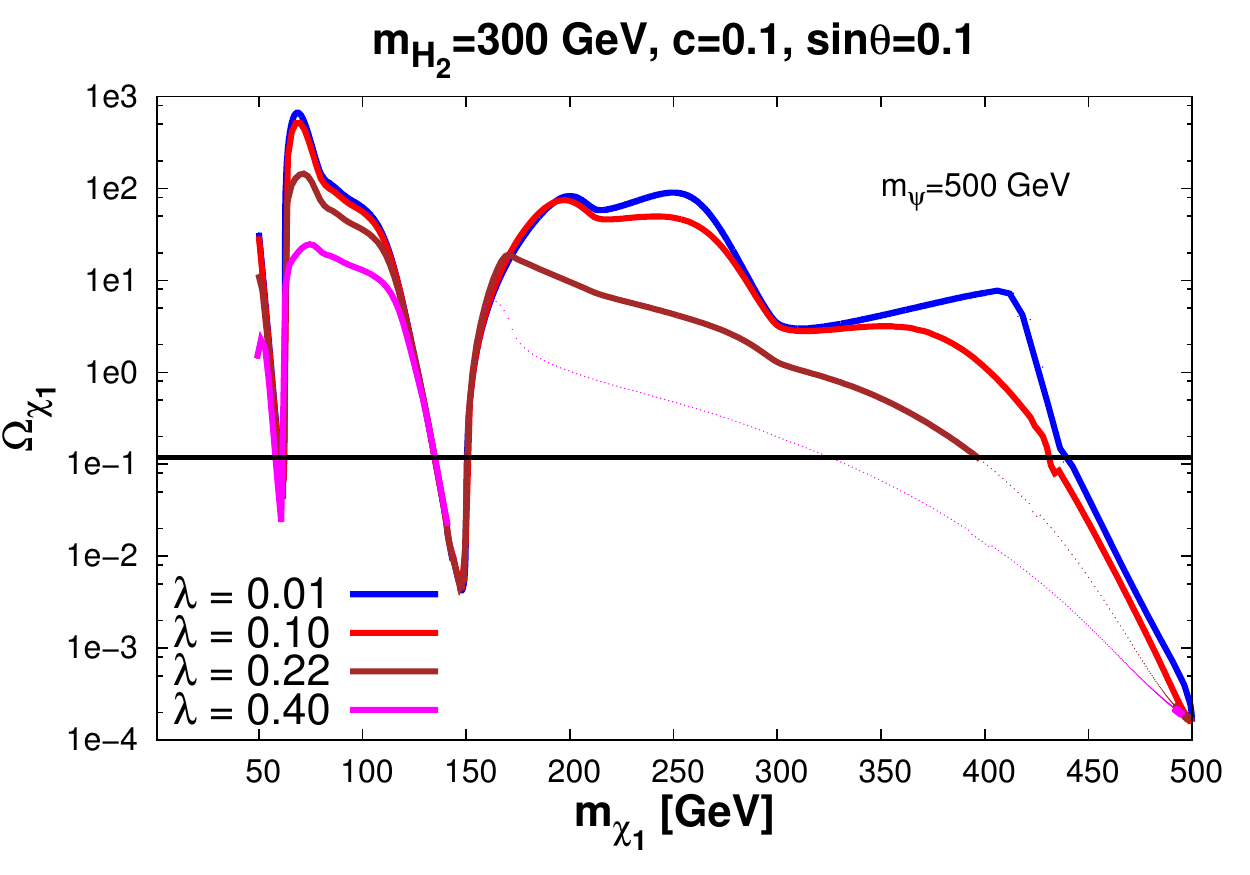}~~
 \includegraphics[height=7 cm, width=7 cm,angle=0]{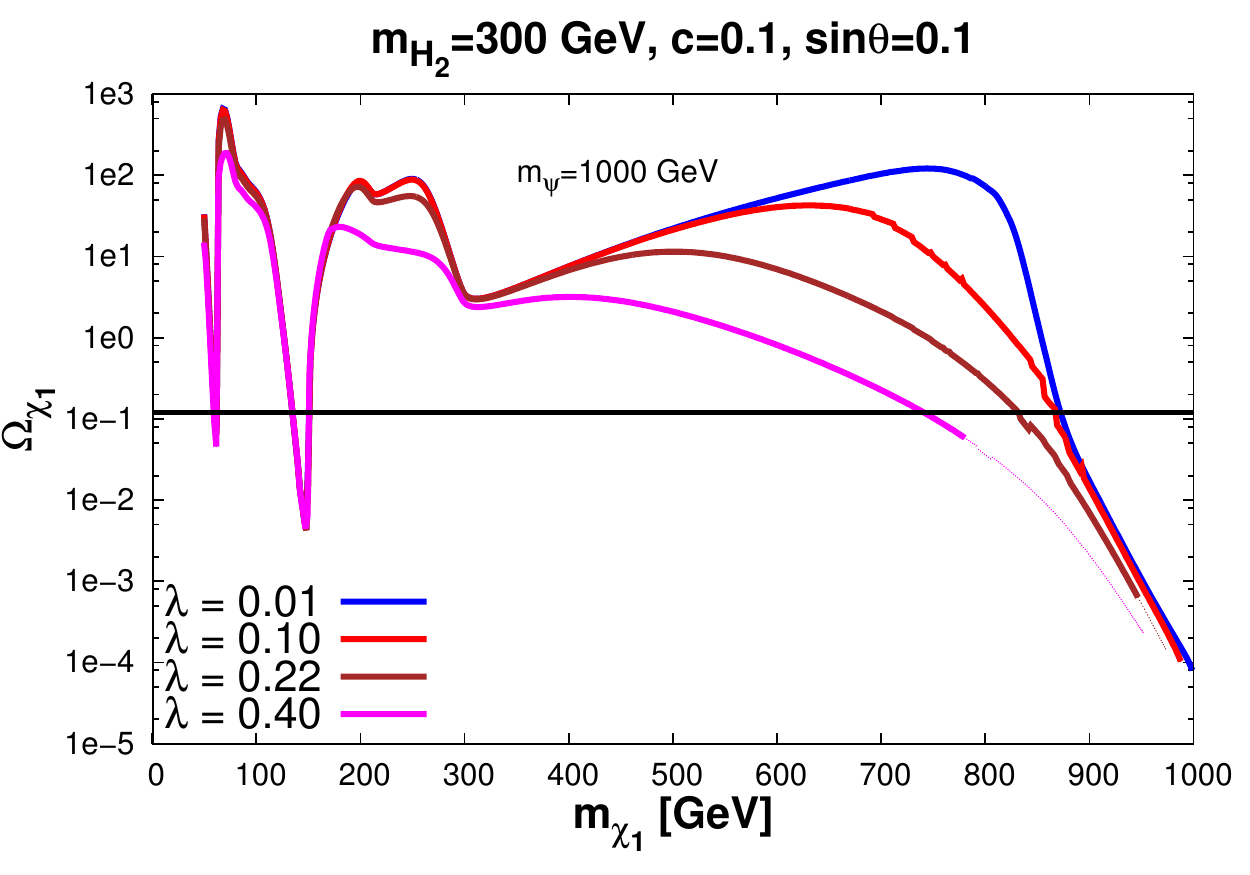}
 \caption{DM relic density as a function of DM mass for [left panel:] $m_{\psi}=$ 500 GeV
 and [right panel:] $m_{\psi}=$ 1000 GeV 
 with  different choices of $\lambda=$ 0.01 (blue), 0.1 (red), 0.25 (brown) and 0.4 (pink).
 Values of heavy Higgs mass, scalar mixing angle and $c$ have been kept fixed at 
 $m_{H_2}=300$ GeV, $\sin\theta=0.1$ and $c=0.1$. Dotted portions indicate the disallowed part
 from SI direct detection cross-section limit.}
 \label{fig1}
 \end{figure}

{Now we would like to investigate how the relic density and direct detection cross-section depend on different parameters of the set-up. For this purpose,}
in Fig.~\ref{fig1} (left panel) 
we plot the variation of DM mass 
$m_{\chi_1}$ with relic density for four different values of Yukawa coupling $\lambda$ while $m_{\psi}$ is 
taken to be 500 GeV. The vev of the singlet scalar $\phi$ is varied from 500 GeV to 10 TeV. Fig.~\ref{fig1} (right panel)
corresponds to a different $m_{\psi} = 1000$ GeV. Other parameters $m_{{\rm{H}}_2}, \sin\theta$ and 
$c$ are kept fixed at 300 GeV, 0.1 and 0.1 respectively as indicated on top of each figures. {Note that $c=0.1$ is a natural choice from the viewpoint that it remains non-perturbative even at very high scale.}
 The horizontal black lines in both the figures denote the required dark matter relic 
abundance. In producing Fig.~\ref{fig1}, dark matter direct detection limits from both spin independent 
and spin dependent searches are included. The solid (colored) portion of a curve correspond to the range 
of $m_{\chi_1}$ which satisfies the SI direct detection (DD) bounds while the dotted portion exhibits the disallowed range
using DD limits.

From Fig.~\ref{fig1}, we also observe that apart from the two resonances, one for the SM Higgs and other 
for the heavy Higgs, the dark matter candidate satisfies the required relic density in another region with 
large value of $m_{\chi_1}$. For example, with $\lambda = 0.22$, the relic 
density and DD cross-section is marginally 
satisfied by $m_{\chi_1} \sim$ 400 GeV. The presence of this allowed value of dark matter mass is due to 
the fact that the co-annihilation processes turn on (they become effective when $\Delta_{i1}/m_{\chi_1} 
\sim 0.1$ or less) which increases the effective annihilation cross-section $\langle \sigma |v| \rangle$ and 
hence a sharp fall in relic density results. Since both annihilation and co-annihilations are proportional to 
$\lambda$ (see Eq.(\ref{eq7})), an increase in $\lambda$ (from pink to red lines) leads to decrease in relic 
density (for a fixed dark matter mass) and this would correspond to smaller value of $m_{\chi_1}$ for 
the satisfaction of the relic density apart from resonance regions. For example, with $\lambda = 0.1$ or 0.01,
 relic density and DD satisfied value of $m_{\chi_1}$ is shifted to $\sim$ 440 GeV compared to $m_{\chi_1} 
 \sim 400$ GeV with $\lambda = 0.22$.  
It can also be traced that there exist couple of small drops of relic density near $m_{\chi_1} \sim 212$ GeV and 
300 GeV. This is mostly prominent for the line with small $\lambda$ (=0.1 (red line) and 0.01 (blue line)). While 
the first drop indicates the opening of the final states $H_1 H_2$, the next one is due to the appearance of 
$H_2 H_2$ final states.

\begin{figure}[H]
 \centering
 \includegraphics[height=7 cm, width=7 cm,angle=0]{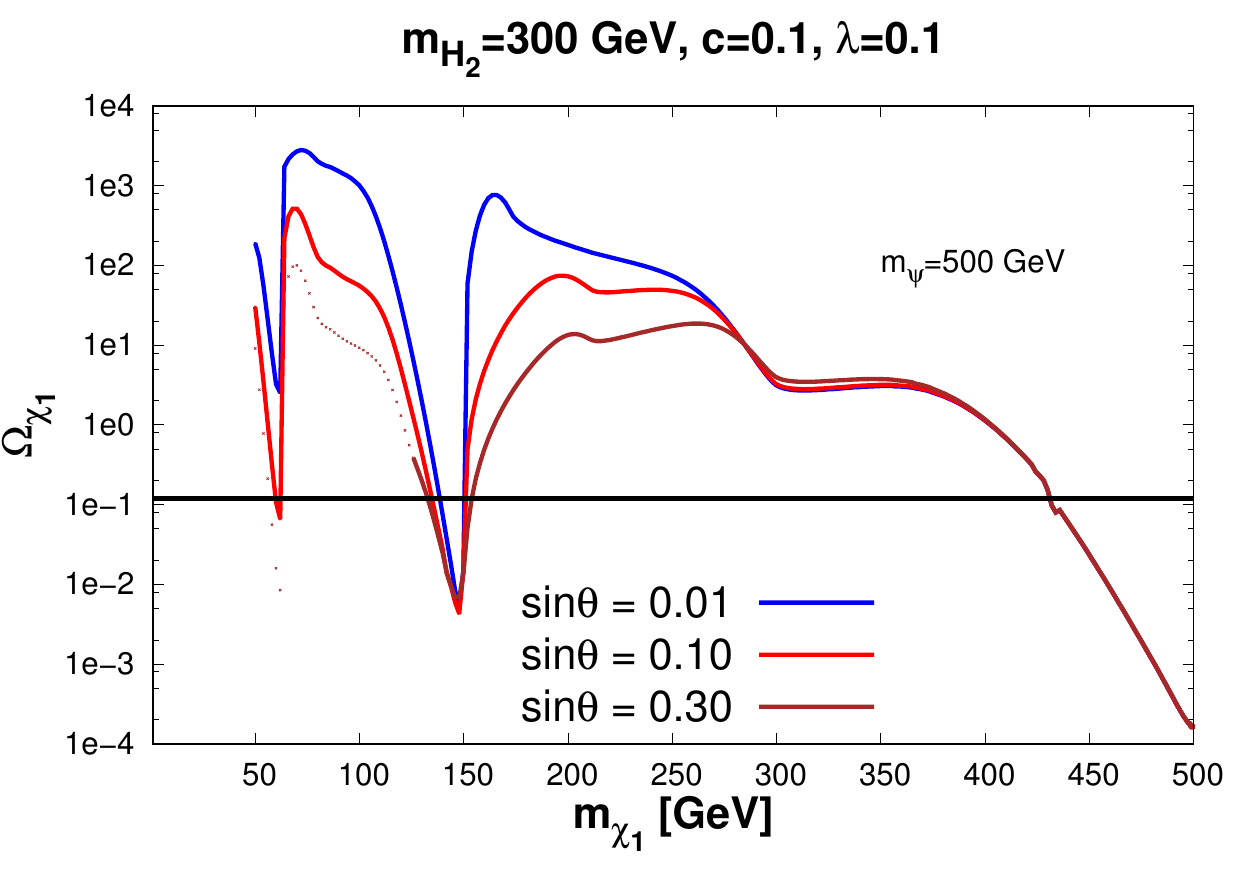}~~
 \includegraphics[height=7 cm, width=7 cm,angle=0]{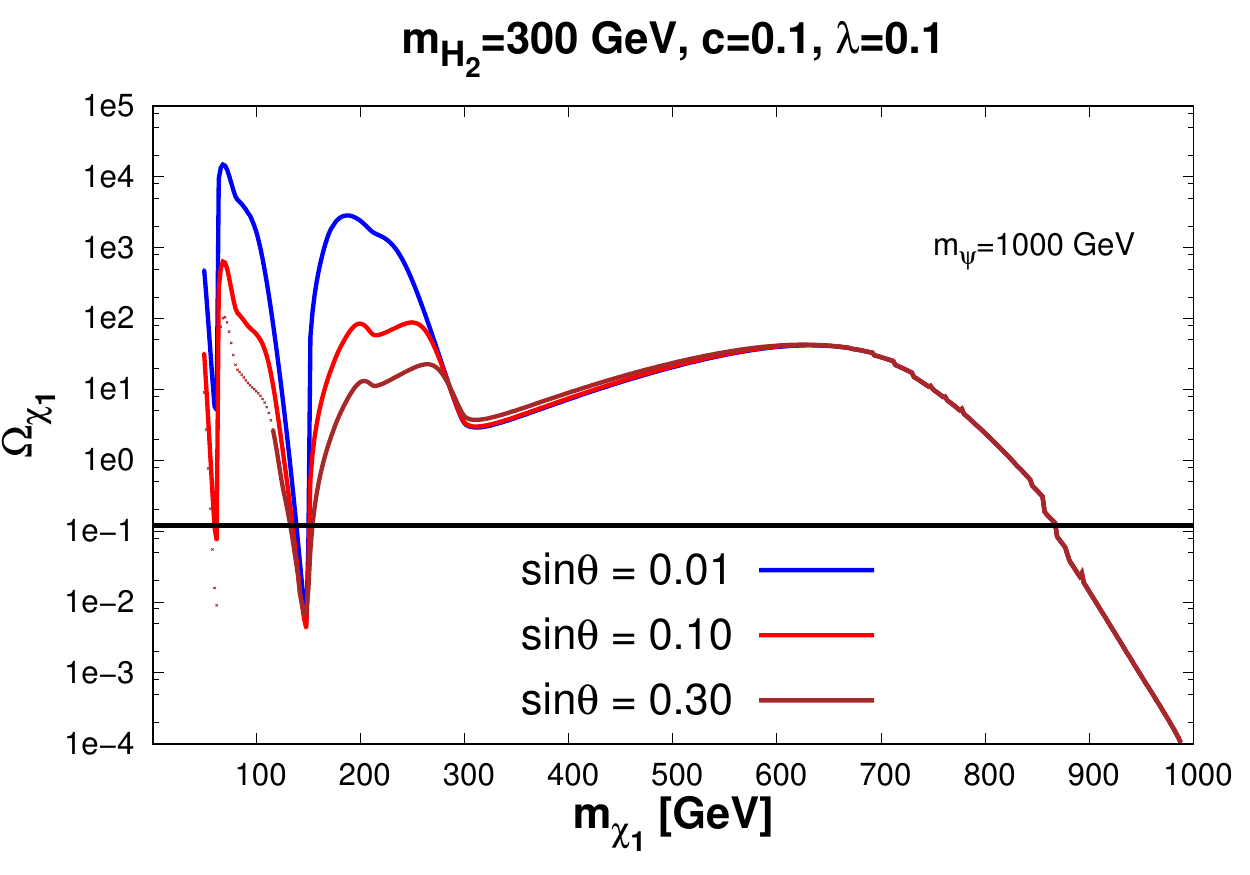}
 \caption{DM relic density as a function of  DM mass considering [left panel:] 
 $m_{\psi}=$ 500 GeV and [right panel:] $m_{\psi}=$ 1000 GeV
 for different choices of scalar mixing angle $\sin\theta\sim$ 0.01 (blue), 0.1 (red) and 0.3 (brown).
 Values of other parameters have been fixed at 
 $m_{H_2}=300$ GeV, $\lambda=0.1$ and $c=0.1$. Dotted portions indicate the disallowed part
 from SI direct detection cross-section limit.}
 \label{fig2}
 \end{figure}
 In the present dark matter model, we found that  
the regions that satisfy dark matter relic density has spin 
dependent cross-section $\sim 10^{-42}-10^{-44}$ cm$^2$ which is well below the present limit obtained 
from spin dependent bounds (for the specific mass range of dark matter we are interested in) from direct 
search experiments \cite{Fu:2016ega}. Therefore, it turns out that the spin independent scattering of dark 
matter candidate is mostly applicable in restricting the parameter space of the present model. 

In Fig.~\ref{fig2}, we depict the effect of scalar mixing in dark matter phenomenology 
keeping parameters $c$ and $\lambda$ both fixed at 0.1 along with the same values of 
$m_{\psi}$ and $m_{H_2}$ used in Fig.~\ref{fig1}. The vev $v_{\phi}$ is varied within 
the range $500~{\rm GeV}\le v_{\phi}\leq10~{\rm TeV}$. Similar to Fig.~\ref{fig1} (there 
with $\lambda$), here also we notice a scaling with respect to different values of $\sin\theta$ as the 
dark matter annihilations depend upon it and there exist two resonances. However 
beyond  $m_{\chi_1} \sim 250$ GeV, dependence on $\sin\theta$ mostly disappears 
as seen from the Fig.~\ref{fig2} as we observe all three lines merge into a single one. Note that 
this is also the region where co-annihilations start to become effective 
as explained in the context of Fig.~\ref{fig1}. It turns out that due to the presence of 
axial type of coupling in the Lagrangian (see Eq.(\ref{eq7})), the co-annihilation processes 
with final state particles including $W^{\pm}$ and $Z$ bosons are most significant and 
they are independent of the scalar mixing $\theta$. It therefore explains the behavior 
of the  red (with $\sin\theta = 0.01$), green (with $\sin\theta = 0.1$) and blue (with 
$\sin\theta = 0.2$) lines in Fig.~\ref{fig2}.

\begin{figure}[H]
\centering
 \includegraphics[height=7 cm, width=7 cm,angle=0]{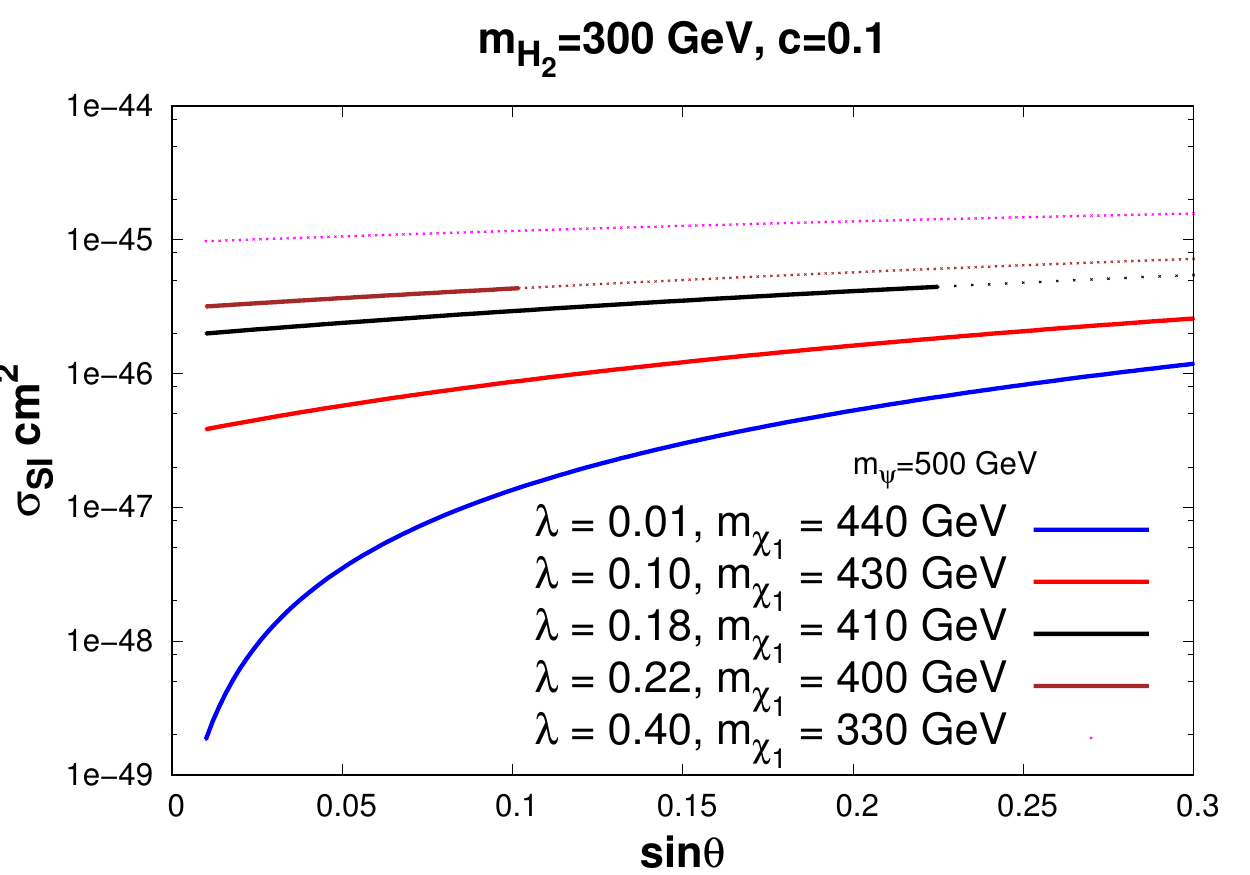}~~
 \includegraphics[height=7 cm, width=7 cm,angle=0]{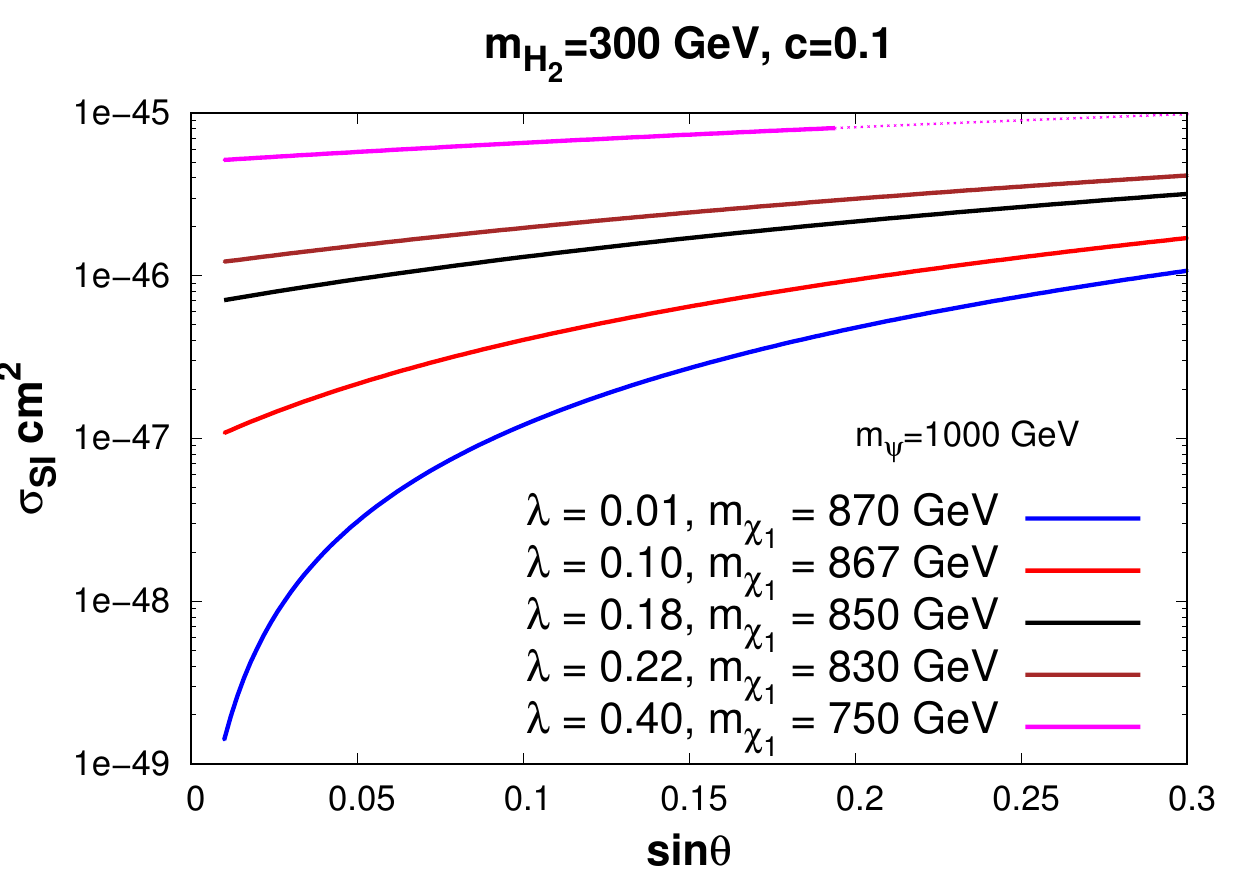}
 \caption{Effects of scalar mixing angle for different values of $\lambda\sim$ 0.01 (blue), 0.1 (red), 0.18 (black),
 0.22 (brown), 0.4 (pink) on dark
 matter spin independent direct detection cross-section for [left panel:] $m_\psi=500$ GeV
 and [right panel:] $m_\psi=1000$ GeV. Values of other parameters have been fixed at 
 $m_{H_2}=300$ GeV and $c=0.1$. Dotted portions indicate the disallowed part
 from SI direct detection cross-section limit.  }
 \label{fig3}
 \end{figure}
From Fig.~\ref{fig2}, it is observed that scalar mixing has not much role to play in the 
co-annihilation region. However the scalar mixing has significant 
effect in the direct detection (DD) of dark matter. To investigate the impact of 
$\sin\theta$ on DD cross-section of DM, we 
choose few benchmark points (set of $\lambda, m_{\chi_1}$ values) in our model that satisfy DM 
relic density excluding the resonance regions ($m_{\chi_1}\simeq m_{H_1}/2$ resonance regime is highly
constrained from invisible Higgs decay limits from LHC).
{Here we vary scalar mixing from 0.01 to 0.3.}

 In Fig.~\ref{fig3}, we show the variation of spin independent dark matter direct 
detection 
cross-section ($\sigma_{SI}$) against $\sin\theta$ for those chosen benchmark values of dark matter mass. 
Keeping parameters $c$ and $m_{H_2}$ fixed at values 0.1 and 300 GeV respectively, $m_{\psi}$
is considered at 500 GeV for the 
left panel and at 1000 GeV for the right panel of Fig.~\ref{fig3}.
  Among these five benchmark sets, four of them (except $\lambda=0.18$, $m_{\chi_1}=410$ GeV)
  were already present in the of Fig.~\ref{fig1} (corresponding to $\sin\theta=0.1$).
 Five lines (blue, red, black, brown and pink colored ones corresponding to different sets of values of $\lambda$ and 
$m_{\chi_1}$) describe the DD cross-section dependence with $\sin\theta$. It is interesting to observe that 
with higher $\lambda$, there exists an increasing dotted portion on the curves (\textit{e.g.} in brown colored line for 
$m_{\psi}=500$ GeV, it starts from $\sin\theta > 0.1$, which stands for the non-satisfaction of the parameter space by the DD limits. This behaviour can be understood in the following way.
 From Eq.(\ref{dd}), it is clear that the first term dominates and hence an increase of SI DD cross-section with respect to 
larger $\sin\theta$ value (keeping other parameters fixed) is expected as 
also evident in the
\begin{figure}[H]
 \centering
\includegraphics[height=7 cm, width=7 cm,angle=0]{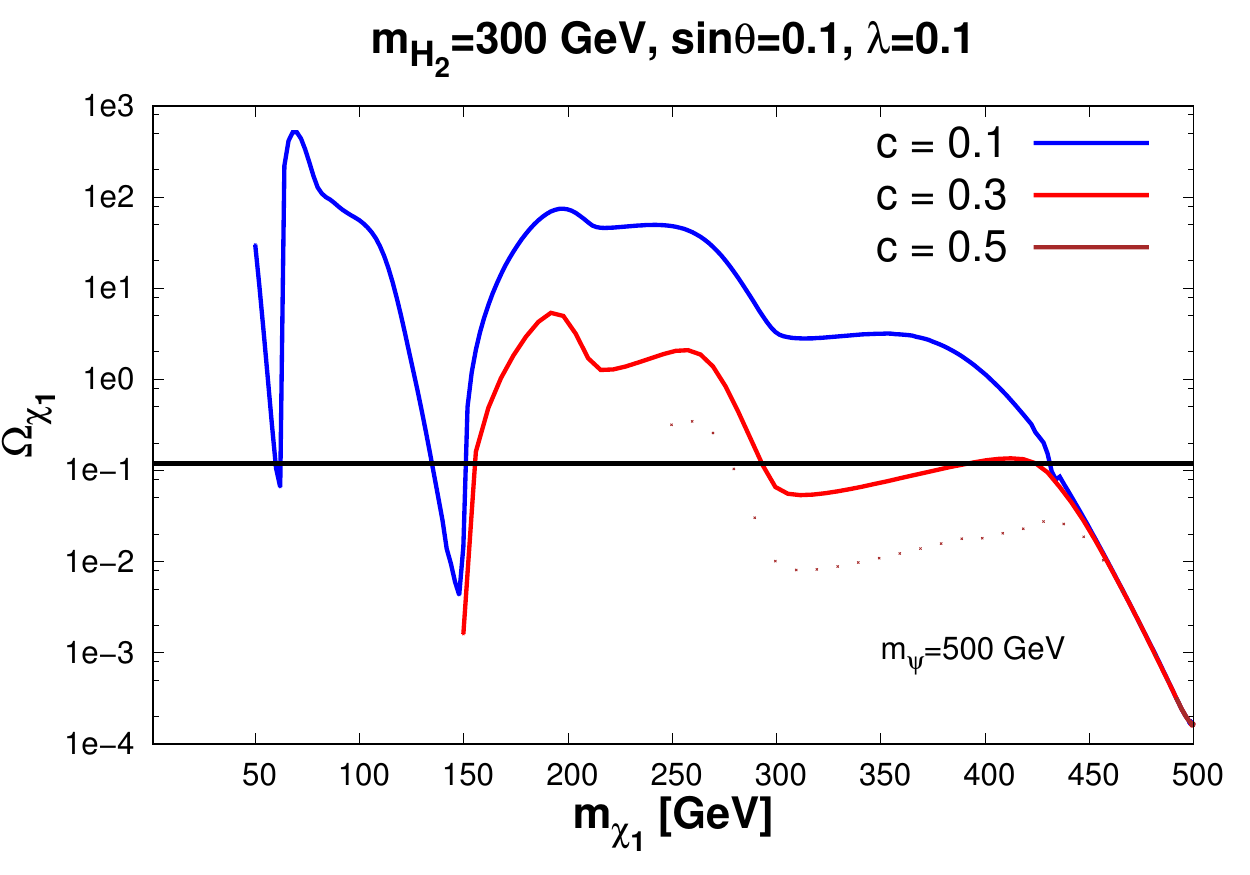}~~
 \includegraphics[height=7 cm, width=7 cm,angle=0]{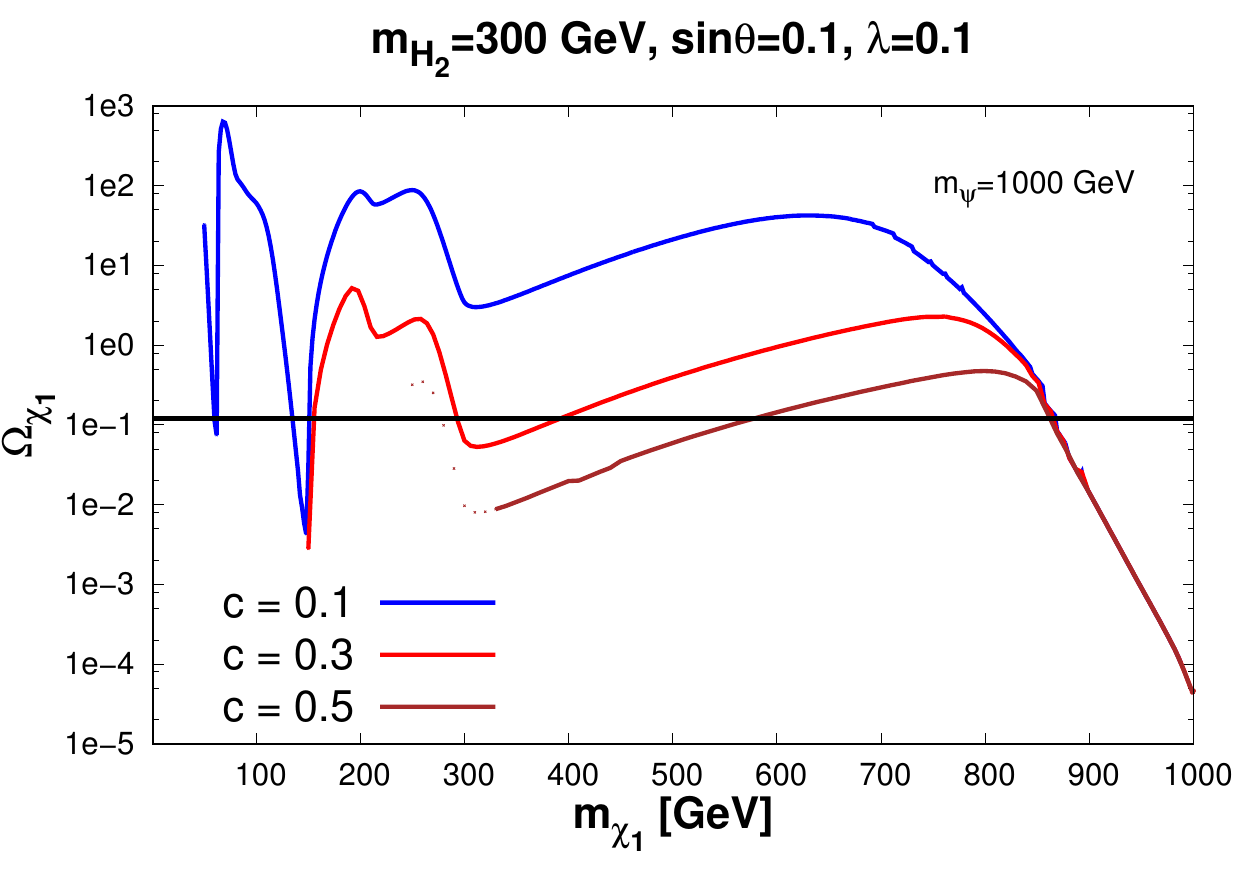}~~\\
 \includegraphics[height=7 cm, width=7 cm,angle=0]{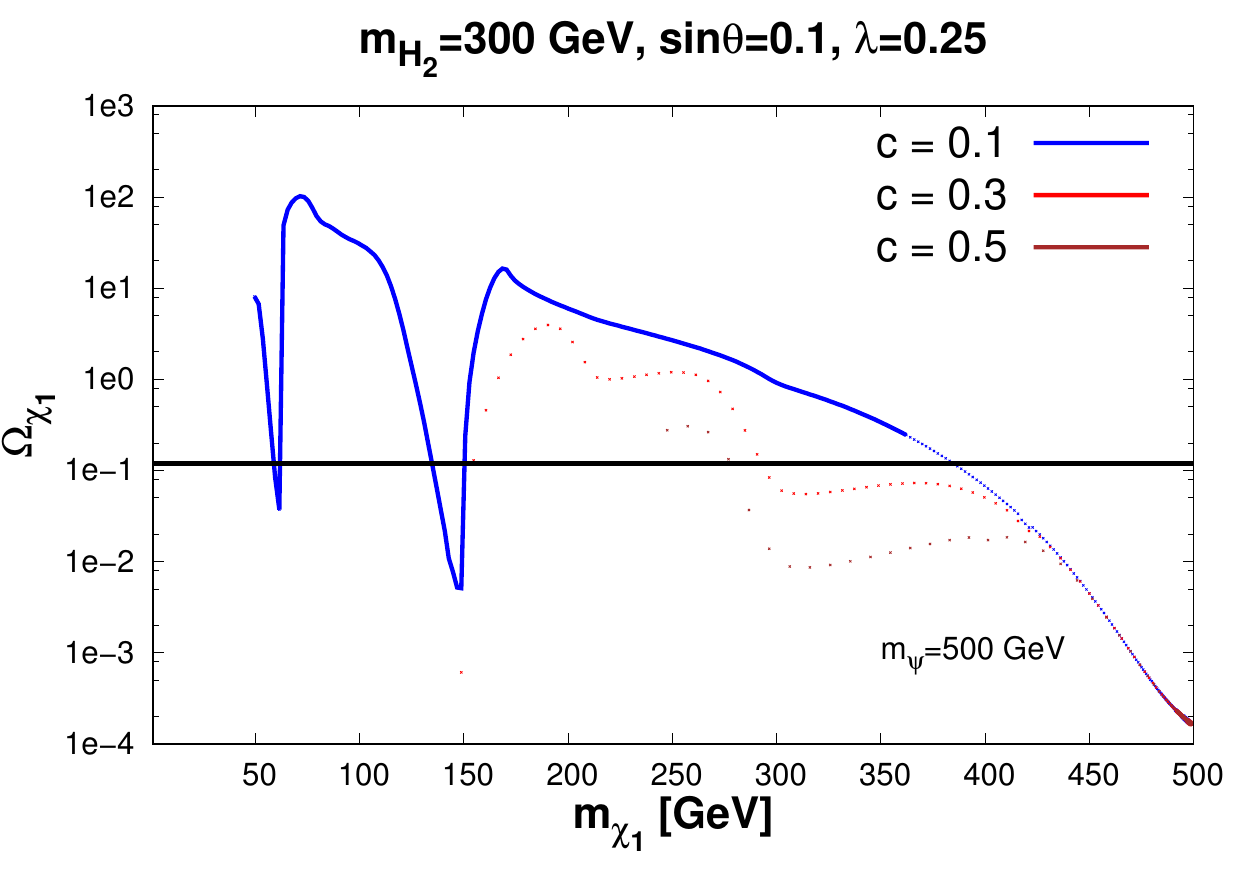}~~
 \includegraphics[height=7 cm, width=7 cm,angle=0]{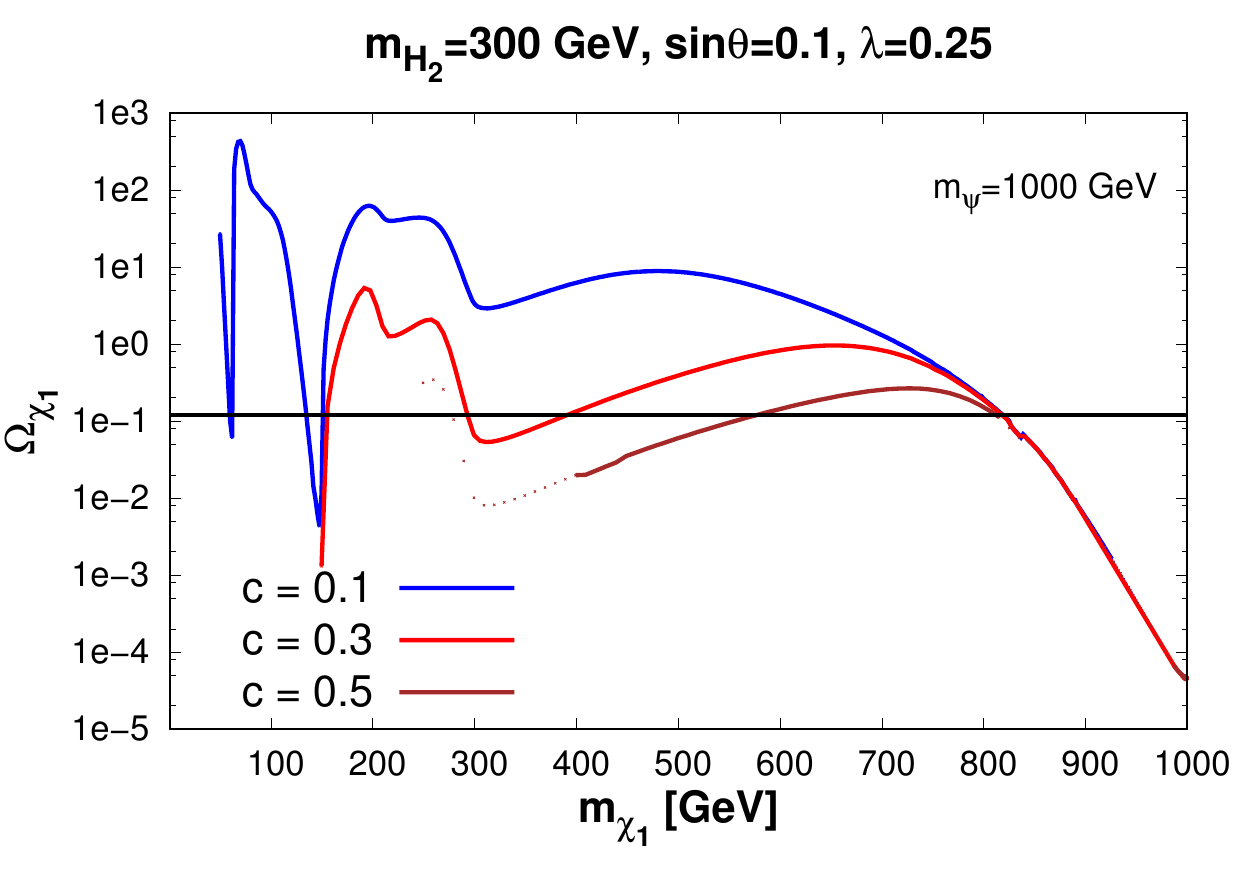}~~
 \caption{Dark matter relic density as a function of DM mass with different choices of
 $c\sim$ 0.1 (blue), 0.3 (red), 0.5 (brown) for [top left:] $m_\psi=500$ GeV, $\lambda=0.1$,
 [top right:] $m_\psi=1000$ GeV, $\lambda=0.1$, [bottom left:] $m_\psi=500$ GeV, $\lambda=0.0.25$ and 
 [bottom right:] $m_\psi=1000$ GeV, $\lambda=0.25$. Values of other parameters have been kept fixed at 
 $m_{H_2}=300$ GeV and $\sin\theta=0.3$. Dotted portions indicate the disallowed part
 from SI direct detection cross-section limit. }
 \label{eff-c}
 \end{figure}

\noindent figures.
We do not include the spin dependent cross-section here; however checked that it remains well within the observed limits.

In Fig.~\ref{eff-c}, we plot the dark matter relic density against
dark matter mass for different values of $c$ keeping other parameters fixed and 
using the same range of $v_{\phi}$ (500 GeV - 10 TeV) as considered in earlier plots. 
The top (bottom) left panel of Fig.~\ref{eff-c} 
corresponds to $m_{\psi}=500$ GeV and top (bottom) right panel are plotted for $m_{\psi}=1000$ GeV.
Curves with higher value of $c$ start with larger initial
value of dark matter mass. 
This can be understood easily from mass matrix $\mathcal{M}$ of Eq.(\ref{eq4}),
as large
\begin{figure}[H]
 \centering
 \includegraphics[height=7 cm, width=7 cm,angle=0]{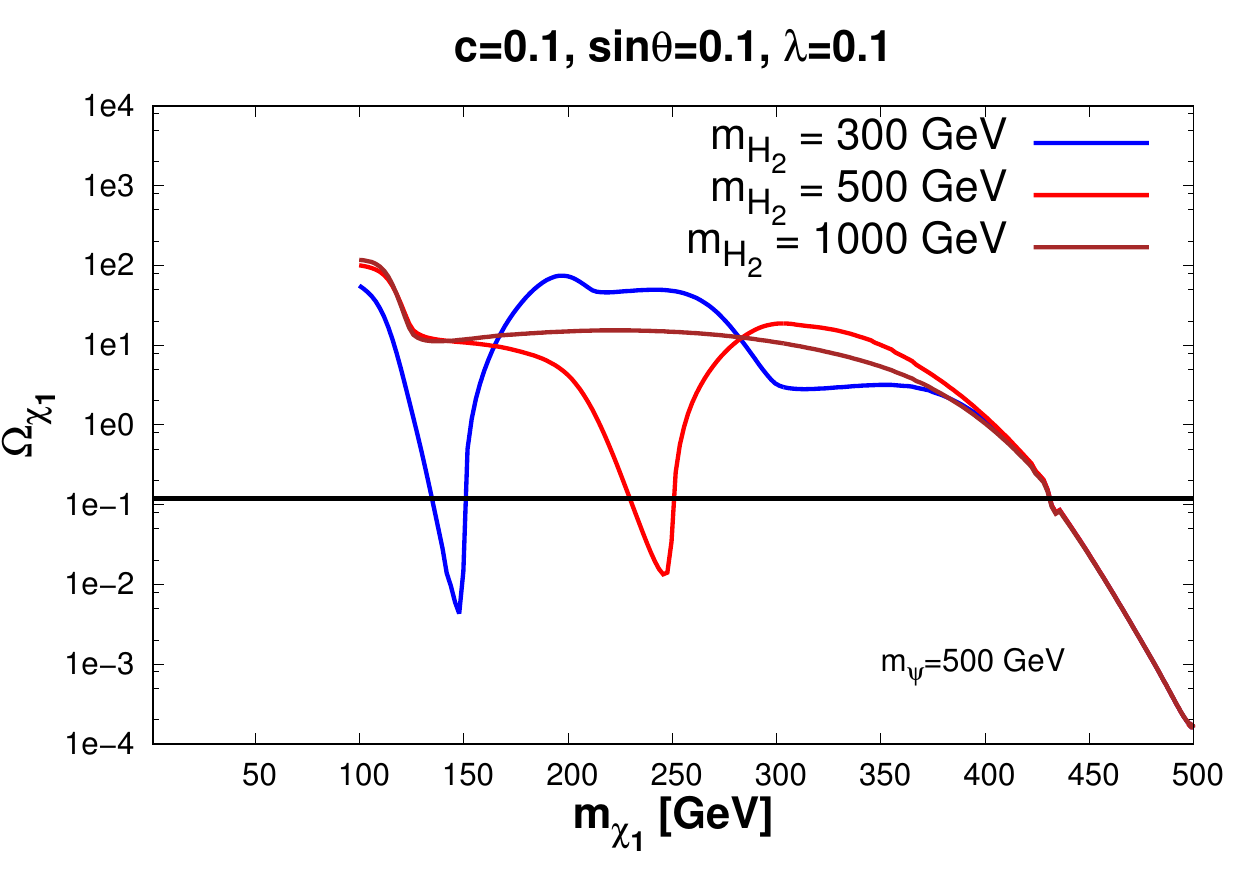}~~
 \includegraphics[height=7 cm, width=7 cm,angle=0]{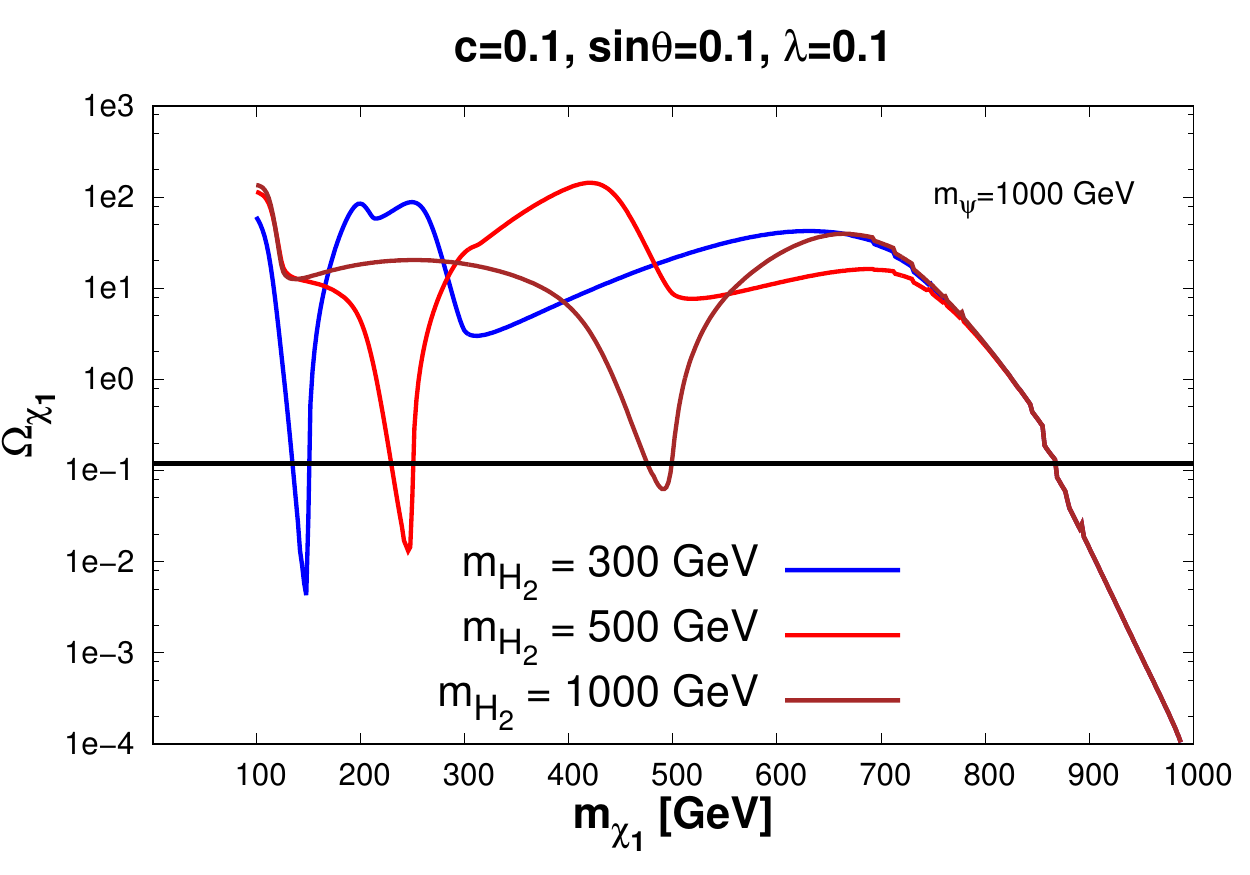}
 \caption{Relic density of dark matter as a function of  DM mass for different
 values of $m_{H_2}=300$ GeV (blue), 500 GeV (red) and 1000 GeV
 (brown) with [left panel:] $m_{\psi}=$ 500 GeV and [right panel:] 
 $m_{\psi}=$ 1000 GeV. The other parameters $c$, $\lambda$ and $\sin\theta$ are kept fixed at 0.1.}
 \label{H2}
 \end{figure}

\noindent  $m_{\psi}$ with small $v_{\phi}$ and $\lambda$, 
dark matter mass $\sim c v_{\phi}$. Hence as $c$ increases, the DM mass starts from a higher value.
The upper panel of figures is for $\lambda = 0.1$ and the lower panel stands for $\lambda = 0.25$.

We observe from Fig.~\ref{eff-c} that enhancing $c$ reduces DM relic density particularly for the region where DM 
annihilation processes are important. At some stage co-annihilation, in particular, processes with 
final states including SM gauge fields takes over which is mostly insensitive to $c$. Hence all 
different curves join together. This is in line with observation in Fig.~\ref{fig2} as well. Here also 

 \noindent we notice that 
all the curves have fall around 212 GeV and 300 GeV where DM DM $\rightarrow H_1, H_2$ and 
DM DM $\rightarrow H_2, H_2$ channels open up respectively. We observe that with a higher value of $c$, for example
with $ c=0.5$ in Fig.~\ref{eff-c} (top left panel), the $DM DM \rightarrow H_2 H_2$ 
annihilation becomes too large and {also} disallowed by the DD 
bounds as indicated by dotted lines. 
We therefore infer that the satisfaction of the DD bounds and the DM relic density prefer a lower value of $c$ which 
is also consistent with the perturbativity point of view. 
Increasing the Yukawa coupling $\lambda$ will change the above scenario, as depicted 
in lower panel of Fig.~\ref{eff-c}. We found that such effect is prominent for smaller values of  $m_{\psi}$ while compared
top and bottom left panels of Fig.~\ref{eff-c}.
 
So far, in Figs.~\ref{fig1}-\ref{eff-c}, we have presented the variations of DM relic
density with DM mass keeping the mass of heavy scalar $H_2$ fixed. 
In Fig.~\ref{H2} (left panel), we show the variation of DM relic 
density against $m_{\chi_1}$ for three different values of $m_{H_2}=300,500,1000$ GeV
with fixed values of $c,~\lambda,~\sin\theta$ (all set to the value 0.1) with $m_{\psi}=500$ GeV. 
 The vev $v_{\phi}$ is varied from 1 TeV to 10 TeV.
 From Fig.~\ref{H2} (left panel), we note that 
each plot for a specific $m_{H_2}$ follow the same pattern as in previous figures. 
Here we notice that with different $m_{H_2}$, the heavy Higgs 
resonance place ($m_{\chi_1}\sim m_{H_2}/2)$ 
\begin{figure}[H]
 \centering
 \includegraphics[height=6 cm, width=8 cm,angle=0]{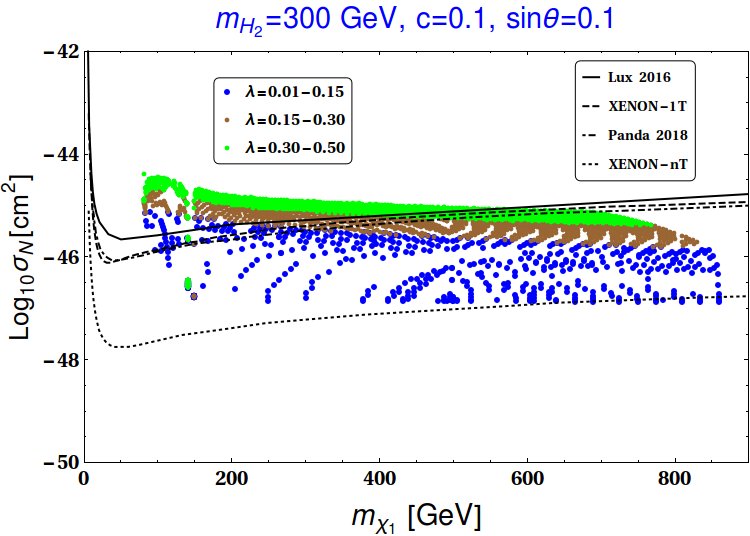}
 \includegraphics[height=6 cm, width=8 cm,angle=0]{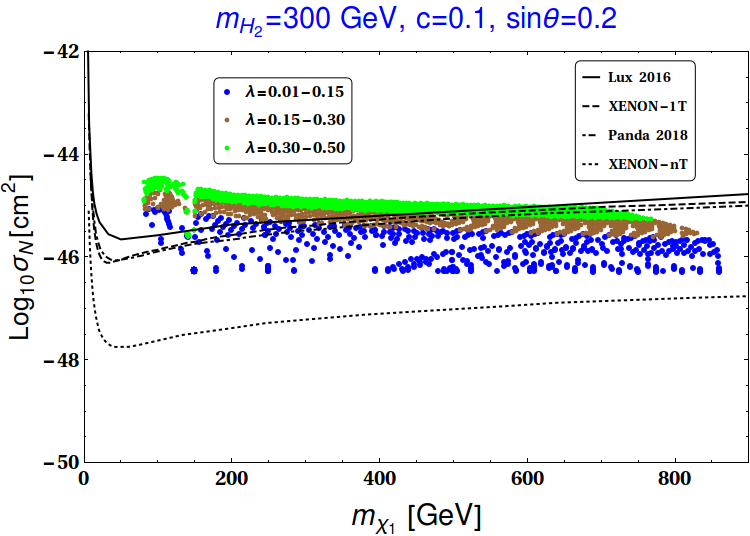}\\
  \includegraphics[height=6 cm, width=8cm,angle=0]{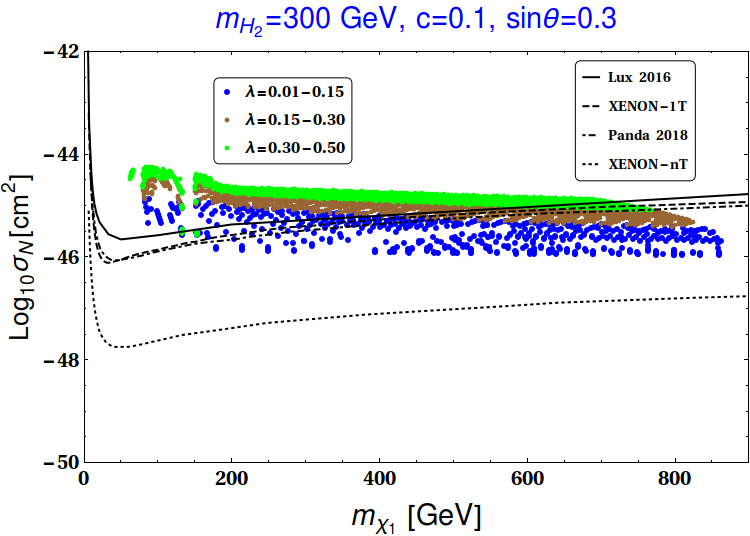}
 \caption{SI direct detection cross-section is plotted against DM mass for 
relic density satisfied points for
 [top left:] $\sin\theta=0.1$, [top right:] $\sin\theta=0.2$ and [bottom:] $\sin\theta=0.3$. The other parameters
 $c=0.1$ and $m_{H_2}=300$ GeV have been kept fixed. Bounds
 from LUX 2016, XENON 1T, PANDA 2018, XENON-nT are also included in the plot.}
 \label{fig4}
 \end{figure}
\noindent is only affected. For large $m_{H_2}$ (say for 1000 GeV), 
the resonance point disappears as it falls within the co-annihilation dominated region. A similar 
plot using the same set of parameters and value of $m_{H_2}$ but with $m_{\psi}=1000$ GeV 
in Fig.~\ref{H2} (right panel) clearly shows this.  In this case, we have a prominent resonance region 
for $m_{H_2}=1000$ GeV as the co-annihilation takes place at a higher value of dark matter mass 
with the increase in $m_{\psi}$.
It is to be mentioned once again that in all the above plots (Figs.~\ref{fig1}-\ref{H2}), 
solid regions indicate the satisfied region
and dotted region indicates the disallowed region for DM mass by spin 
independent direct detection cross-section bounds.

\subsubsection{Constraining $\lambda-\sin\theta$ from a combined scan of parameters}
A more general result for the present dark matter model can be obtained by varying the 
mass of charged fermion $m_{\psi}$, vev $v_{\phi}$ of the heavy singlet scalar field 
and the Yukawa coupling $\lambda$. We use the LEP bound on chargino mass to
set the lower limit on the mass of charged fermion $m_{\psi}\gtrsim 100$ GeV \cite{Abdallah:2003xe}.
Using this limit on charged fermion mass, we scan the parameter space of the model 
with the following set of parameters
\bea
100~{\rm GeV} \lesssim m_{\psi} \lesssim 1000~{\rm GeV};~ 500~{\rm GeV}\lesssim v_{\phi} \lesssim 10~{\rm TeV};~
0.01 \lesssim \lambda \lesssim 0.5;\nonumber\\
\textrm{               \hspace{-2cm}          }
\sin\theta~=~0.1,0.2,0.3;~
c=0.1;~~m_{H_2}=300~{\rm GeV.}
\label{range}
\eea

In Fig.~\ref{fig4} (top left panel)
we plot the values of DM mass against dark matter spin independent cross-section 
for the above mentioned ranges of parameters with $\sin\theta=0.1$ which  already 
satisfy DM relic abundance obtained from Planck \cite{Ade:2015xua}. 
Different ranges of the Yukawa coupling $\lambda$ are shown 
in blue (0.01-0.15), brown (0.15-0.30) and green (0.30-0.50) shaded regions.

The bounds on DM mass and SI direct detection scattering cross-section
from LUX\cite{Akerib:2016vxi}, XENON-1T\cite{Aprile:2017iyp}, Panda 2018\cite{Cui:2017nnn} 
and XENON-nT\cite{Aprile:2015uzo} are also shown for comparison. The spin dependent scattering 
cross-section for the allowed parameter space is found to be in agreement with the present limits from Panda 2018\cite{Cui:2017nnn} 
and does not provide any new constraint on the present phenomenology.
From Fig.~\ref{fig4} (top left panel) it can also be observed that
increasing $\lambda$ reduces the region allowed by the most stringent Panda 2018 limit. This is 
due to the fact that an increase in $\lambda$ enhances the dark matter direct detection 
cross-section as we have clearly seen from previous plots (see Fig. \ref{fig1}). Here we observe that with the 
specified set of parameters, dark matter with mass above 100 GeV is consistent with 
DD limits with $\lambda = 0.01-0.15$ (see the blue shaded region). For the brown region, 
we conclude that with $\lambda = 0.15-0.30$, DM mass above 400 GeV is allowed and 
with high $\lambda = 0.30-0.50$, DM with mass 600 GeV or more is only allowed. 
We also note that a large region of the allowed parameter space is ruled out when XENON-nT \cite{Aprile:2015uzo}
direct detection limit is taken into account.

Similar plots for the same range of parameters given in Eq.(\ref{range})
for $\sin\theta=0.2$ and 0.3 are shown in top right panel and bottom panel of Fig.~\ref{fig4} respectively. These plots 
depict the same nature as observed in top left panel of Fig.~\ref{fig4}. In all these plots, 
the low mass region ($m_{\chi_1}\lesssim 62.5$ GeV) is excluded due to invisible decay bounds on Higgs and $Z$.  
It can be observed comparing all three plots in Fig.~\ref{fig4}, that the allowed region 
of DM satisfying relic density and DD limits by Panda 2018 becomes shortened with the increase 
$\sin\theta$. In other words, it prefers a larger value of DM mass with the increase of $\sin\theta$.
This is also expected as the increase of $\sin\theta$ is associated with larger 
DD cross-section (due to $H_1, H_2$ mediated diagram).  Hence overall we conclude from this 
DM phenomenology that increase of both $\lambda$ and $\sin\theta$ push the allowed value 
of DM mass toward a high value.  In terms of vacuum stability, these two parameters, the 
Yukawa coupling $\lambda$ and the scalar mixing $\sin\theta$, affect the Higgs vacuum stability 
differently. The Yukawa coupling destabilizes
the Higgs vacuum while the scalar mixing $\sin\theta$ makes the vacuum more stable. Detailed 
discussion on the Higgs vacuum stability is presented in the next section.
\begin{figure}[H]
 \centering
 \includegraphics[height=7cm, width=10cm,angle=0]{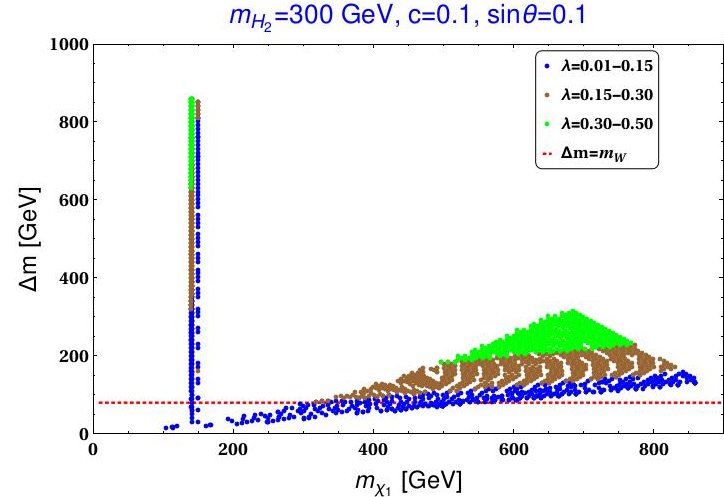}\\
 
 \vspace{-3mm}
 \caption{Mass difference between DM and the charged fermion $\Delta m$ is plotted
 against $m_{\chi_1}$ for different sets of $\lambda\sim$ 0.01-0.15 (blue), 0.15-0.30 (brown) 
 and 0.30-0.50 (green) with $\sin\theta = 0.1$.
  All points satisfy the relic density direct detection cross-section bound 
from PANDA 2018.
  The other parameters are kept fixed at $c=0.1$ and $m_{H_2}=300$ GeV. The red line indicates the $W$ boson mass ($m_W$).}
 \label{Deltam}
 \end{figure}
A general feature of the singlet doublet model is the existence of two other neutral fermions, $\chi_{2,3}$ 
and a charged fermion, $\psi^{+}$. All these participate in the co-annihilation process which 
contributes to the relic density of the dark matter candidate, $\chi_1$. 
{The charged fermion $\psi^+$ can decay into 
$W^+$ and $\chi_1$, when the mass splitting $\Delta m= m_{\psi}-m_{\chi_1}$
is larger than $W^+$ mass. However, for 
mass splitting $\Delta m$} between $\chi_1$ and $\psi^+$ 
smaller than the mass of gauge boson $W^+$, the three body decay of charged 
fermion, $\psi^+$ into ${\chi}_1$ 
associated with lepton and neutrino becomes plausible. This 
three body decay must occur before 
${\chi}_1$ freezes out, otherwise it would contribute to the relic. Therefore, the decay lifetime of 
$\psi^+$ should be smaller compared to the freeze out time of ${\chi}_1$. The freeze out of the 
dark matter candidate ${\chi}_1$ takes place at temperature $T_f=m_{\chi_1}/20$.
Therefore, the corresponding freeze out time can be expressed as 
\be
t=1.508 g_{\star}^{-\frac{1}{2}} M_P/T_f^2 \,\, ,
\ee
where $M_P$ is the reduced Planck mass $M_P=2.435\times10^{18}$ GeV and $g_{\star}$ is effective number 
of degrees of freedom.
The decay lifetime of the charged fermion $\psi^+$ is given as $\tau_{\psi^+}=\frac{1}{\Gamma_{\psi^+}}$,
where $\Gamma_{\psi^+}$ is the decay width for the decay $\psi^+ \rightarrow {\chi}_1 l^+ \bar{\nu}_l$,
is of the form
\bea
\Gamma_{\psi^+}&=&\frac{G_F^2}{12\pi^3} \left[(V_{31}^2+V_{21}^2)\{-2m_{\psi}m_{\chi_1}^2 I_1 +
3(m_{\psi}^2+m_{\chi_1}^2)I_2-4m_{\psi}I_3\}
\right . \nonumber \\&& \left .
+12V_{31}V_{21}\{m_{\chi_1}(m_{\psi}^2+m_{\chi_1}^2)I_1-2m_{\psi}m_{\chi_1}I_2\}\right] \,\, .
\label{decay}
\eea
 \begin{figure}[h]
 \centering
 \includegraphics[height=6 cm, width=8 cm,angle=0]{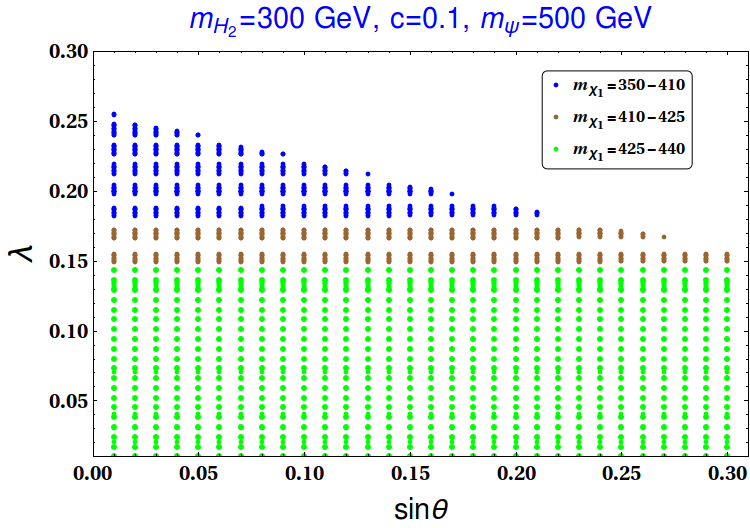}
 \includegraphics[height=6 cm, width=8 cm,angle=0]{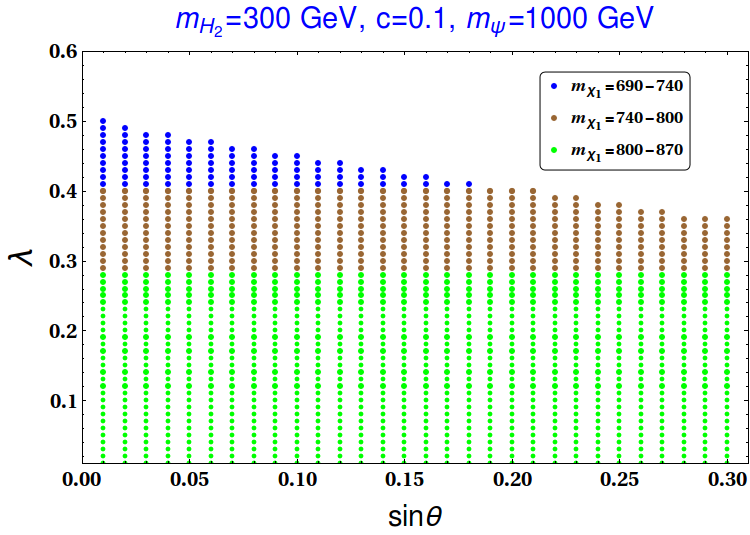}
 \caption{Correlation between $\lambda$ and $\sin\theta$ for both relic  and 
direct detection cross-section
 satisfied points with [left panel:] $m_\psi=500$ GeV and [right panel:] $m_\psi=1000$ GeV. The other parameters
 are kept fixed at $c=0.1$, $m_{H_2}=300$ GeV. Different ranges of $m_{\chi_1}$ (in GeV) are shown with color codes
 as mentioned in the inset.}
 \label{lambda-sin}
 \end{figure}

 \noindent In the above expression, $G_F$ is the Fermi constant and the terms $I_{1,2,3}$ is expressed as 
\bea
I_1=\int\sqrt{x^2-a^2}dx,~~I_2=\int x \sqrt{x^2-a^2}dx,~~I_3=\int x^2 \sqrt{x^2-a^2}dx \,\, 
\label{integral}
\eea
where $x=E_{\chi_1}$ and $a=m_{\chi_1}$, $E_{\chi_1}$ being the total energy of $\chi_1$.

In order to satisfy the condition that $\psi^+$ decays before the freeze out of ${\chi}_1$, one must
have $\tau_{\psi_+} \le t$. The integrals $I_{1,2,3}$ in Eq.(\ref{integral}) are functions of mass splitting 
$\Delta m$ and so is the total decay width $\Gamma_{\psi^+}$. To show the dependence on $\Delta m$,
we present a correlation plot $m_{\chi_1}$ against $\Delta m$ in Fig.~\ref{Deltam}. Fig.~\ref{Deltam}
is plotted for the case $\sin\theta=0.1$ (consistent with Fig.~\ref{fig4} having the region
allowed by the DD bound from Panda 2018). We use the same color code
for $\lambda$ as shown in Fig.~\ref{fig4}. The horizontal red line
indicates the the region where $\Delta m=m_{W}$. From Fig.~\ref{Deltam} we observe that for smaller 
values of $\lambda$ (0.01-0.15), $\Delta m<m_W$ is satisfied upto $m_{\chi_1}\sim 500$ GeV. The mass
splitting increases for larger $\lambda$ values.
We find that for the chosen range of model parameters (Eq.(\ref{range})), the decay life time
$\tau_{\psi^+}$ is several order of magnitudes smaller than the freeze out time of $\chi_1$.

We end this section by estimating the value of $T$ parameter in Table~\ref{tab:TP} for two sets of relic 
satisfied points (with $\lambda= 0.4$ and 0.18) as we mentioned before that among
the $S,T$ and $U$, only $T$ would be relevant 
in this scenario. 
\begin{table}[h]
\begin{center}
\begin{tabular}{ |c|c|c|c|c|c| } 
\hline
 $c$ & $m_\psi (GeV)$  & $v_\phi$ (TeV) & $m_{\chi_1} $ (GeV) & $\lambda$ & T$\times 10^{-4}$\\
\hline 
 0.1 & 1000 & 7.55 & 750  & $0.4$ & $1$ \\ 
\hline
0.1 & 500 & 4.20 & 410  & $0.28$ & $0.1$ \\  
 \hline
\end{tabular}
\end{center}
\caption{Values of T-parameter induced by extra fermions in the set up for two sets of relic density satisfied points
(see Fig.~\ref{fig3}).}
\label{tab:TP}
\end{table}
With further smaller $\lambda$, $T$ parameter comes out to be very small and hence it does not 
pose any stringent constraint on the relic satisfied parameter space.
However with large $\lambda\sim 1$, the situation may alter.

 In our scenario, we have also seen in Fig.~\ref{fig1} that 
for value of $\lambda$ larger than 0.4, the direct detection cross-section of dark matter 
candidate also increases significantly and are thereby excluded by present limits on dark matter 
direct detection cross-section. 
To make this clear, here we present a plot, Fig. \ref{lambda-sin} (left panel), of relic density and DD satisfied 
points in the $\sin\theta-\lambda$ plane, where the other parameters are fixed at $c =0.1, m_{\psi} = 500$ GeV, 
$m_{H_2} = 300$ GeV. As before, $v_{\phi}$ is varied between 500 GeV and 10 TeV. 
Similar plot with same set of 
$c,~m_{H_2}$ but with $m_{\psi}=1000$ GeV is depicted in right panel of Fig.~\ref{lambda-sin}.
Different ranges of dark matter masses are specified with different colors as mentioned in the caption of 
Fig.~\ref{lambda-sin}. From Fig.~\ref{lambda-sin} we observe that allowed range of $\lambda$ reduces
 with the increase of scalar mixing due to the DD bounds. From Fig.~\ref{lambda-sin} (left panel) we get a
maximum allowed $\lambda \sim 0.25$ while the same for the $m_{\psi}=1000$ GeV (right panel) turns out to be $\lambda =0.5$.
Furthermore as we will see the study of vacuum stability, discussed  in Sec.~\ref{vs}, indicates that the 
Yukawa coupling $\lambda$ should not be large in order to maintain the electroweak vacuum absolutely stable till 
Planck scale. Therefore, larger values of $\lambda$ (close to 1) is not favoured in the present scenario.

\section{EW vacuum stability}
\label{vs}
In the present work consisting of singlet doublet dark matter model with additional scalar, we have already 
analysed (in previous section) the parameter space of the set-up using the relic density and direct 
detection bounds. Here we extend the analysis by examining the Higgs vacuum stability within the 
framework. It is particularly interesting as the framework contains two important parameters, (i)
coupling of dark sector fermions with SM Higgs doublet ($\lambda$) and (ii) mixing (parametrized by angle $\theta$) 
between the singlet scalar and SM Higgs doublet. The presence of these two will modify the stability of the 
EW vacuum. First one makes the situation worse than in the SM by 
{driving} the Higgs quartic coupling 
$\lambda_H$ negative earlier than $\Lambda_I^{\textrm{SM}}$. 
The second one, if sufficiently large, can negate 
the effect of first and make the Higgs vacuum stable. Thus the stability of Higgs vacuum depends on 
the interplay between these two. Moreover, as we have seen, the scalar singlet also enriches the 
dark sector with several new interactions that significantly contribute to DM phenomenology satisfying 
the observed relic abundance and direct detection constraint. Also the scalar mixing angle is bounded 
by experimental constraints ($\sin\theta\lesssim 0.3$) as we have discussed in Sec.~\ref{constraints}. 

 The proposed set up has two additional mass scales: the DM mass ($m_{\chi_1}$) and heavy Higgs ($m_{H_2}$). 
 Although the dark sector has four physical fermions (three neutral and one charged), we can safely ignore 
 the mass differences between them when we consider our DM to fall outside the 
 two resonance regions (Figs. \ref{fig1}-\ref{H2}). As we have seen 
 in this region (see Fig.~\ref{Deltam}), co-annihilation becomes dominant, all the masses in the dark sector 
 fermions are close enough ($\sim m_{\chi_1}$, see Figs.~\ref{fig1}-\ref{H2}). Hence the 
 renormalisation group (RG) equations will be modified accordingly from SM ones with the relevant couplings 
 entering at different mass scales. Here we combine the  RG equations (for the relevant couplings only) \cite{Lu:2017uur}
together in the following (provided $\mu>m_\phi,m_{\chi_1}$),
\begin{align}
\frac{dg_1}{dt}&=\beta^{\textrm{SM}}_{g_1}+\frac{1}{16 \pi^2}\frac{2}{3}g_1^3,\\
\frac{dg_2}{dt}&=\beta^{\textrm{SM}}_{g_2}+\frac{1}{16 \pi^2}\frac{2}{3}g_2^3,\\
\frac{d\lambda_H}{dt}&=\beta^{\textrm{SM}}_{\lambda_H}+\frac{1}{16 \pi^2}\Big\{\frac{\lambda_{\phi H}^2}{2}\Big\}+\frac{1}{16 \pi^2}\Big\{-2 \lambda^4 + 4\lambda_H \lambda\Big\}\label{eq:lambda},\\
\frac{dy_t}{dt}&=\beta^{\textrm{SM}}_{y_t}+\frac{1}{16 \pi^2}\Big\{\lambda^2 y_t\Big\},\\     
\frac{d\lambda}{dt}&=\frac{1}{16 \pi^2} \Big\{\lambda (3 y_t^2 - \frac{3}{4} g_1^2- \frac{9}{4} g_2^2) + 
   \frac{5}{2} \lambda^3\Big\},\\
 \frac{d\lambda_{\phi H}}{dt}&=\frac{1}{16\pi^2}\Big\{12\lambda_H\lambda_{\phi H}+6\lambda_\phi\lambda_{\phi H}+4\lambda_{\phi H}^2+6 y_t^2
 \lambda_{\phi H}-\frac{3}{2}g_1^2\lambda_{\phi H}-\frac{9}{2}g_2^2\lambda_{\phi H}+2\lambda^2\lambda_{\phi H}+2 c^2 \lambda_{\phi H}\Big\},\nonumber\\
 \frac{d\lambda_{\phi}}{dt}&=\frac{1}{16\pi^2}\Big\{18\lambda_\phi^2+2\lambda_{\phi H}^2-\frac{1}{2}c^4 + 4 \lambda_\phi c^2\Big\},\\
\frac{d c}{dt}&=\frac{1}{16\pi^2}\Big\{6 c^3\Big\},
\end{align}
where $\beta^{\textrm{SM}}$ is the SM $\beta$ function (in three loop) of respective 
couplings \cite{Buttazzo:2013uya,Mihaila:2012fm,Mihaila:2012bt,Bednyakov:2012en}.

In this section our aim is to see whether we can achieve SM Higgs vacuum stability till Planck mass ($M_P$).
However we have two scalars (SM Higgs doublet and one gauge singlet $\phi$) in the model. Therefore we should ensure the
boundedness or stability of the entire scalar potential in any field direction. In that case the following matrix
\begin{align}
\begin{pmatrix}
    \lambda_H      & \frac{\lambda_{\phi H}}{2} \\
     \frac{\lambda_{\phi H}}{2}      & \lambda_\phi \\
\end{pmatrix},
\end{align}
has to be co-positive. The conditions of co-positivity \cite{Kannike:2012pe,Chakrabortty:2013mha} of such a 
matrix is provided by
\begin{align}
 \lambda_{H}(\mu)>0,\textrm{ }\lambda_{\phi}(\mu)>0, \textrm{ and }\lambda_{\phi H}(\mu)+2\sqrt{\lambda_H(\mu)\lambda_\phi(\mu)
 }>0.
 \label{coposi}
\end{align}
Violation of $\lambda_H>0$ could lead to unbounded potential or existence of another deeper minimum 
along the Higgs direction. 
The second condition ($\lambda_{\phi}(\mu)>0$) restricts the scalar potential from having any runway 
direction along $\phi$. Finally,
$\lambda_{\phi H}(\mu)+2\sqrt{\lambda_H(\mu)\lambda_\phi(\mu)}>0$ ensures the potential to be bounded from below 
or non-existence of another deeper minimum somewhere between $\phi$ or $H$ direction.

On the other hand, if there exists another deeper minimum other than the EW one, the estimate of the
tunneling probability $P_T$ of the EW vacuum to the second minimum is essential. The Universe will be in metastable
state only, provided the decay time of the EW vacuum is longer than the age of the Universe. The tunneling
probability is given by \cite{Buttazzo:2013uya,Isidori:2001bm},
\begin{align}
 P_T=T_U^4\mu_B^4e^{-\frac{8\pi^2}{3|\lambda_H(\mu_B)|}}.
\end{align}
where $T_U$ is the age of the Universe, $\mu_B$ is the scale at which probability is maximized, determined
from $\beta_{\lambda_H} = 0$. Hence metastable Universe requires \cite{Buttazzo:2013uya,Isidori:2001bm}
\begin{align}
 \lambda_H(\mu_B) ~>~ \frac{-0.065}{1-\textrm{ln}\Big(\frac{v}{\mu_B}\Big)}.
\label{eq:Met}
\end{align}
As noted in \cite{Buttazzo:2013uya}, for $\mu_B > M_P$, one can safely consider $\lambda_H(\mu_B)=\lambda_H(M_P)$.

The RG improved effective Higgs potential (at high energies $H_0\gg v$) can be written as \cite{Casas:1994qy,Degrassi:2012ry}
\begin{align}
V_H^{\textrm{eff}}=\frac{\lambda_H^{\textrm{eff}}(\mu)}{4}H_0^4,
\end{align}
with $\lambda_H^{\textrm{eff}}(\mu)=\lambda_H^{\textrm{SM, eff}}(\mu)+\lambda_H^{\phi,\textrm{eff}}(\mu)
+\lambda_H^{(\psi_{D_1},\psi_S,)\textrm{eff}}(\mu)$ where 
 $\lambda_H^{\textrm{SM, eff}}$ is the Standard Model contribution to $\lambda_H$. The other two contributions
 $\lambda_H^{\phi,\textrm{eff}}$ and $\lambda^{(\psi_{D_1},\chi),\textrm{eff}}$ are due to the newly added fields 
in the present model as provided below.  
  \begin{align}
&\lambda_H^{\phi,\textrm{eff}}(\mu)=e^{4\Gamma(H_0=\mu)}\Big[\frac{\lambda_{\phi H}^2}{64\pi^2}\Big(\textrm{ln}
\frac{\lambda_{\phi H}}{2}-
\frac{3}{2}\Big)\Big],\\
&\lambda_H^{(\psi_{D_1},\chi),\textrm{eff}}(\mu)=e^{4\Gamma(H_0=\mu)}\Big[\frac{\lambda^4}
{16\pi^2}\Big(\textrm{ln}\frac{\lambda}{2}-\frac{3}{2}\Big)\Big].
\end{align}
Here $\Gamma(H_0)=\int_{m_t}^{H_0}\gamma(\mu)d\textrm{ln}\mu$, $\gamma(\mu)$ is the anomalous dimension
of the Higgs field\cite{Buttazzo:2013uya}.

In SM, the top quark Yukawa coupling ($y_t$) drives the Higgs quartic coupling to negative values.
 In our set up, the coupling $\lambda$ has very similar effect on $\lambda_H$ in Eq.(\ref{eq:lambda}). 
 So combination of both $y_t$ and $\lambda$ make the situation worse (by driving Higgs vacuum more towards 
 instability) than in SM.
 However due to the presence of extra singlet scalar, $\lambda_H$ gets a positive threshold
 shift (second term in Eq.(\ref{lambdaH}) at energy scale $m_{H_2}$. Also the RG equation of $\lambda_H$
 is aided by a positive contribution from the interaction of SM Higgs with the extra scalar ($\lambda_{\phi H}$).
 Here we study whether these two together can negate the combined effect of $y_t$ and $\lambda$ 
 leading to $\lambda_H^{\textrm{eff}}>0$ for
all energy scale running from $m_t$ to $M_P$. Note that the threshold shift (second term in Eq.(\ref{lambdaH})) 
in $\lambda_H$ is function of $m_{H_2}$ and $\sin\theta$. On the other hand, other new 
couplings relevant for study of EW vacuum stability are $\lambda_\phi$ and $\lambda_{\phi H}$ which 
can be evaluated from the values of $m_{H_2}$, $v_\phi$ and $\sin\theta$ through Eq.(\ref{lambdaphi}) 
and Eq.(\ref{lambdaChiH}). Hence once we fix $m_{H_2}$ and use the SM values of Higgs and top mass, the 
stability analysis effectively depends on $\lambda$, $\sin\theta$ and $v_\phi$. We run the three loop RG equations for all the
 SM couplings and one loop RG equations for the other relevant couplings in the model 
 from $\mu=m_t$ to $M_P$. We use the boundary conditions of SM couplings as 
provided in Table \ref{tab:ini}. The boundary values have been evaluated in \cite{Buttazzo:2013uya}
by taking various threshold corrections at $m_t$ and mismatch between top pole mass and 
$M\bar{S}$ renormalised couplings into account.
\begin{table}[H]
\begin{center}
\begin{tabular}{|c | c | c | c | c | c |}
\hline
Scale & $y_t$ & $g_1$ & $g_2$ & $g_3$ & $\lambda_H$\\
\hline
$\mu=m_t $ & $0.93610$ & $0.357606$ & $0.648216$ & $1.16655$ & $0.125932$\\
\hline
\end{tabular}
\end{center}
\caption{Values of  the relevant SM couplings (top-quark Yukawa $y_t$, gauge couplings $g_i$ and $\lambda_H$) at
 energy scale $\mu=m_t=173.2$ GeV with $m_h~(m_{H_1})=125.09$ GeV and $\alpha_S(m_Z)=0.1184$.}
\label{tab:ini}
\end{table}

\section{Phenomenological implications from DM analysis and EW vacuum stability}
\label{EWph}
{We have already found the correlation between $\lambda$ and $\sin\theta$ to satisfy the
relic abundance and spin independent DD cross-section limits on DM mass as displayed in
Fig. \ref{lambda-sin}. It clearly shows that for comparatively larger value of $\lambda$, upper limit 
on $\sin\theta$ from DD cross-section is more restrictive. 
On the
other hand, a relatively large value of $\lambda$ affects the
 EW vacuum stability adversely. In this regard, a judicious choice of 
reference points from Fig. \ref{lambda-sin} is made
in fixing benchmark points (BP-I of Table \ref{tab:iniDM}, corresponding to left panel of Fig. \ref{lambda-sin} and BP-II of Table \ref{tab:iniDM} corresponding to right panel of Fig. \ref{lambda-sin}). 
BP-I and BP-II involve moderate values of $\lambda$ 
for which DD-limits starts constraining $\sin\theta$ (more than the existing constraints as per Sec. 3) and $\lambda_H$ gets
significant running. These points would then indeed test the viability of the model. Note that the benchmark
 points (BP-I and II) are also present in Fig. \ref{lambda-sin} in which restrictions on $\sin\theta$
from DD limits are explicitly shown.}
\begin{table}[H]
\begin{center}
\begin{tabular}{|c | c | c | c | c | c | c |}
\hline
Benchmark points & $m_{\chi_1}$ (GeV) & $m_\psi$ (GeV) & $m_{H_2}$ (GeV) & $c$ & $v_\phi$ (TeV) & $\lambda$\\
\hline
BP-I & 410 & 500 & 300 & $0.1$ & 4.2 & 0.18\\
\hline
BP-II & 750 & 1000 & 300 & 0.1 & 7.55 & 0.4\\
\hline
\end{tabular}
\end{center}
\caption{Initial values of the relevant mass scales (DM mass $m_{\chi_1}$ and heavy Higgs mass $m_{H_2}$), $v_\phi$
 and the couplings ($c$ and $\lambda$) of the dark sector used to study the Higgs vacuum stability.}
\label{tab:iniDM}
\end{table}
\begin{figure}[h]
 \centering
 \includegraphics[height=6 cm, width=8 cm,angle=0]{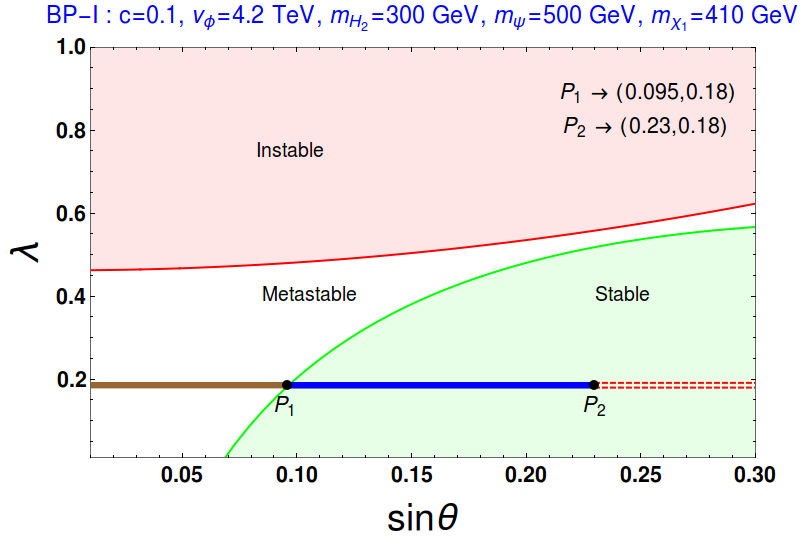}
 ~~\includegraphics[height=6 cm, width=8 cm,angle=0]{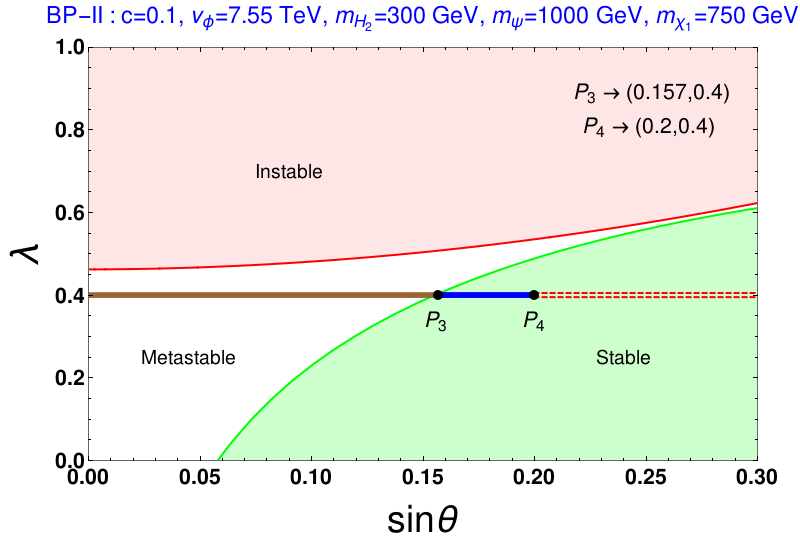}
 \caption{{Vacuum stability (green), metastable (white) and instability (pink) region in $\sin\theta-\lambda$
 plane for  BP-I [left panel] and BP-II [right panel]. The horizontal lines describe correct relic density contours for BP-I and BP-II. The red dashed portion of these horizontal  lines represent the
 disallowed regions from direct detection limit. The blue section of each relic contour satisfies both vacuum stability criteria as well as the direct detection bound while the brown portion is excluded by vacuum
stability condition only. The points $P_1$ and $P_2$ in left panel and $P_3$ and $P_4$ from right panel will be used to show the
evolution of $\lambda_H$ as a function of energy scale $\mu$ in Fig. \ref{fig:lambdaHCoposi}.}}
 \label{fig:SinLam}
  \end{figure}

 In Fig.\ref{fig:SinLam}, we constrain the $\sin\theta-\lambda$ parameter space using the
absolute stability criteria ($\lambda_H^{\textrm{eff}}(\mu)>0$ for $\mu=m_t$ to $M_P$) 
for the EW vacuum for BP-I and BP-II as values of parameters given in Table~\ref{tab:iniDM}.
The solid green line in Fig.~\ref{fig:SinLam} indicates the boundary line in $\sin\theta-\lambda$ plane 
beyond which the  stability criteria of SM Higgs vacuum violates. Hence all points in the green shaded 
region satisfies the absolute stability of the EW vacuum. Similarly the solid red line indicates the boundary 
of the metastable-instable region as obtained through Eq.(\ref{eq:Met}). 
  The pink shaded region therefore {indicates} instability of the EW 
vacuum with $m_t =173.2$ GeV and 
$m_h = 125.09$ GeV. Here we use the upper limit on the scalar mixing as 0.3 so as to be consistent with 
experimental limits on it.
The DD cross-section corresponding to these particular dark matter masses (410 GeV and 750 GeV) 
with specific choices of $\lambda$ ($\lambda =0.18$ in left plot
while $\lambda = 0.4$ in right plot in Fig.~\ref{fig:SinLam}) against 
$\sin\theta$ along with the same values of other parameters ($c, v_{\phi}, m_{\psi}, m_{H_2}$) 
are already provided in Fig. \ref{fig3} {(also in Fig. \ref{lambda-sin})}. Using  Fig.~\ref{fig3}, we identify here the relic density satisfied contour 
(horizontal solid line) in the $\sin\theta-\lambda$ plane on Fig.\ref{fig:SinLam}. We note that DD sets an 
upper bound on $\sin\theta$, due to which the excluded region of $\sin\theta$ is marked in red within 
the horizontal line(s) in both the figures. The brown portion of the $\lambda = 0.18 (0.4)$ line 
corresponds to the relic and DD allowed range of $\sin\theta$ in left(right) plot; however this falls in a 
region where EW vacuum is metastable. In Fig.~\ref{fig:SinLam} (left panel),
the blue portion of the constant $\lambda$ line indicates 
that with this restricted region of $\sin\theta$, we have a dark matter of mass 410 GeV which 
satisfy the relic density and DD bounds and on the other hand, the EW vacuum remains absolutely 
stable all the way till $M_P$.

The outcome of this combined analysis of relic and DD satisfied value of a DM mass and 
stability of the EW vacuum in presence of two new scales, DM mass and heavy Higgs, 
seems to be interesting. It can significantly restrict the scalar mixing angle. For example 
with $m_{\chi_1}=410$ GeV in Fig.~\ref{fig3} (left panel) and $m_{H_2} = 300$ GeV, 
we find $\lambda = 0.18$ restricts $\sin\theta \lesssim 0.23$ which is more stringent than the 
existing experimental one.
{This set of ($\sin\theta,\lambda$) values is denoted by $P_2$ in left panel of 
\begin{figure}[H]
 \centering
\includegraphics[height=5.2 cm, width=7 cm,angle=0]{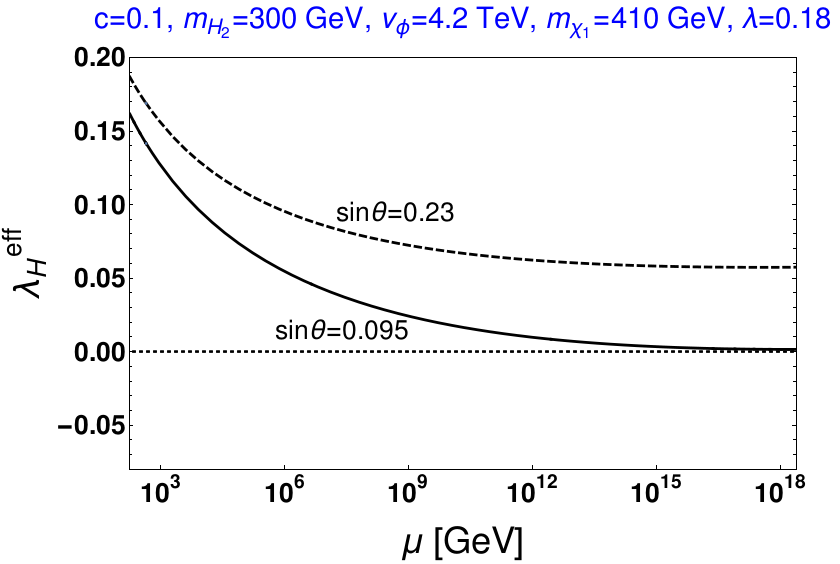}
 ~~~~~\includegraphics[height=5.2 cm, width=7 cm,angle=0]{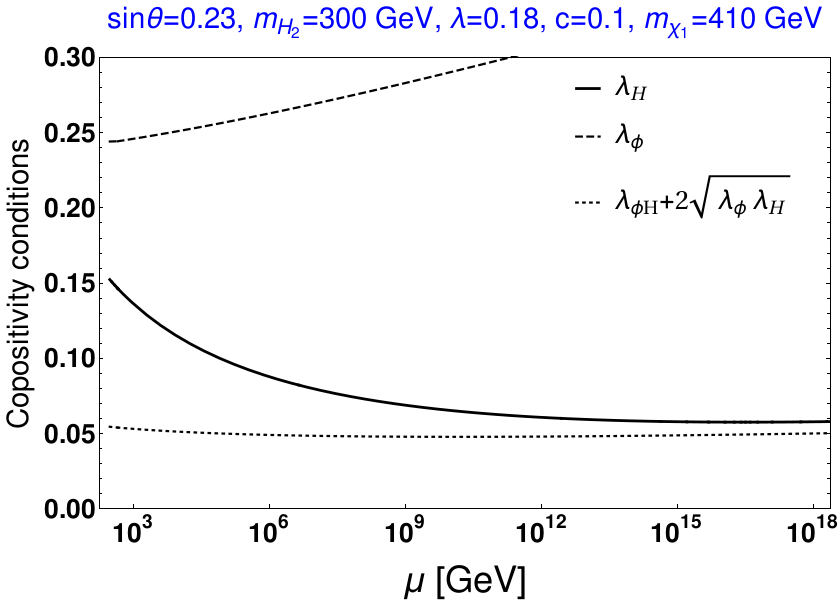}\\
 \includegraphics[height=5.2 cm, width=7 cm,angle=0]{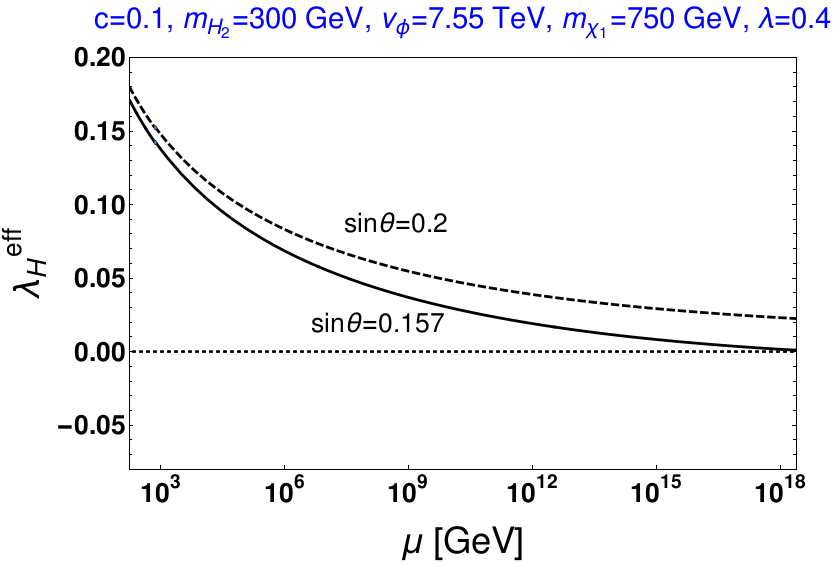}
 ~~~~~\includegraphics[height=5.2 cm, width=7 cm,angle=0]{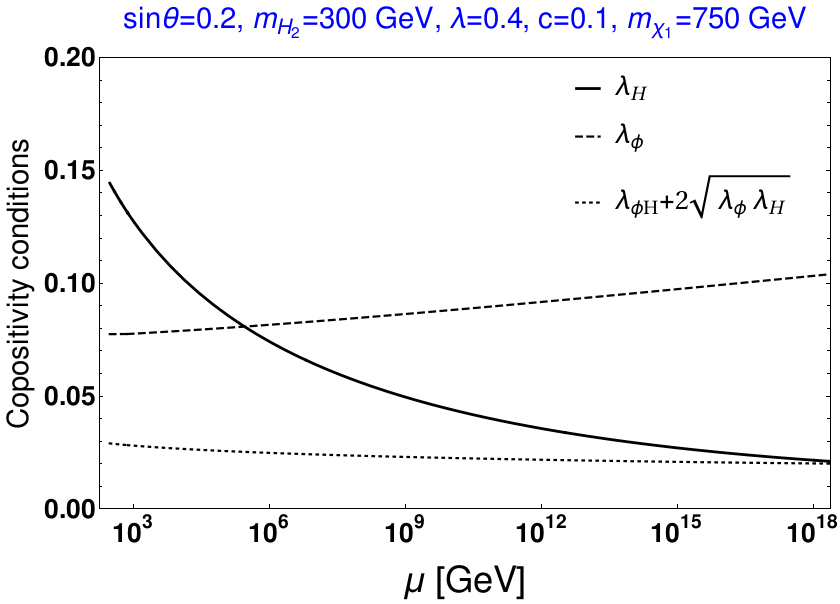}
 \caption{{Evolution of $\lambda_H^{\textrm{eff}}$ from $\mu=m_t$ to $M_P$ for $P_1$, $P_2$ [top left] and  $P_3$, $P_4$ [bottom left] points of
 Fig.\ref{fig:SinLam}. In right panels, copositivity criterias are shown as a function of $\mu$ for $P_2$ [top] and $P_4$ [bottom] points.}}
 \label{fig:lambdaHCoposi}
\end{figure}
\noindent Fig. \ref{lambda-sin}.}  On top of this, if the EW vacuum needs to be absolutely stable, 
we note that we can obtain a lower limit on $\sin\theta$ as 0.095. {The corresponding set of ($\sin\theta,\lambda$) values is denoted by $P_1$ which follows from the intersection of relic density contour ($\lambda=0.18$ line) with boundary 
line of the absolute stability region (solid green line)}. Combining these we obtain: 
$0.095 \lesssim\sin\theta \lesssim 0.23$. A similar criteria with $m_{\chi_1} = 750$ GeV restricts $\sin\theta$ 
to be within {$0.157 ~({\rm point~} P_3) \lesssim \sin\theta \lesssim 0.2 ~({\rm point ~} P_4)$}. {Therefore, from this analysis
we are able to draw both upper and lower limits on $\sin\theta$ 
{for} the two benchmark points. This turns out to be the most 
interesting
and key feature of the
proposed model.
 The vacuum stability analysis can be extended for any other 
points in Fig. \ref{lambda-sin}. However if we go for higher value of $\lambda$, the simultaneous satisfied 
region of DM relic abundance, DD cross-section bound and stability of EW vacuum 
will be reduced as seen while comparing the left with right panel of Fig. \ref{fig:SinLam}.}

\hspace{4mm} {Finally one may wonder about the nature of evolution of $\lambda_H^{\textrm{eff}}$ and the co-positivity 
conditions for any points within EW vacuum stability 
satisfied region of Fig. \ref{fig:SinLam}. Hence, in Fig.\ref{fig:lambdaHCoposi},
running of $\lambda_H^{\textrm{eff}}$ is shown against the energy scale $\mu$ for $P_1$
and $P_2$ (in top left panel of Fig.~\ref{fig:lambdaHCoposi}); $P_3$ and $P_4$ (in bottom left panel of Fig.~\ref{fig:lambdaHCoposi}}). 
Note that these two points also satisfy the relic density and DD cross-section 
bounds. We find for $\sin\theta=0.2$, 
$\lambda_H^{\textrm{eff}}$ remains positive starting from $\mu=m_t$ to $M_P$ energy scale and for 
$\sin\theta=0.157$, although $\lambda_H^{\textrm{eff}}$ stays positive throughout its evolution, it 
marginally reaches zero at $M_P$. Hence this point appears as the boundary point in 
$\sin\theta-\lambda$ plane of Fig.~\ref{fig:SinLam} (right panel) beyond which the SM Higgs vacuum becomes unstable.
{In top and bottom right panel of Fig.~\ref{fig:lambdaHCoposi}, we show the evolution of all the co-positivity conditions 
from $\mu=m_t$ to $M_P$ corresponding to $P_2$ and $P_4$ points respectively.}
\section{Conclusion}
\label{conclusion}

We have explored a dark matter model by extending the Standard Model of particle 
physics with a singlet scalar and a dark sector comprised of two Weyl doublets 
and a Weyl singlet fermions. The scalar singlet acquires a vev and 
contributes to 
the mass the dark sector particles consisting of three neutral Majorana fermions and 
one charged Dirac fermion. The lightest Majorana particle is stable due to the 
presence of a residual $Z_2$ symmetry and hence we study whether this can 
account for the dark matter relic density and also satisfy the direct detection bounds.
There exists a mixing of the singlet scalar with the SM Higgs doublet in the model which 
results in two physical scalars, which in turn affect the DM phenomenology. We have 
found that apart from the region of two resonances, there exists a large available 
region of parameter space satisfying various theoretical and experimental bounds 
particularly due to large co-annihilations effects present. On the other hand, inclusion 
of new fermions in the model affects the Higgs vacuum stability adversely by 
leading it more toward instability at high scale due to new Yukawa like coupling. This issue 
however can be resolved by the involvement of extra scalar singlet. We find that 
with the demand of having a dark matter mass $\sim$ few hundred GeV to 1 TeV consistent 
with appropriate relic density and DD limits and simultaneously to make the EW vacuum 
absolutely stable upto the Planck scale, we can restrict the scalar mixing angle significantly. 
The result is carrying a strong correlation with the dark sector Yukawa coupling, $\lambda$. 
It turns out that, with higher dark matter mass, the allowed range of $\sin\theta$ becomes 
more stringent from this point of view. Hence future limits of $\sin\theta$ will have the potential to allow or rule out the 
model under consideration. 

\vskip 1 cm 

\noindent {\bf Acknowledgments} : ADB and AS acknowledge the support from Department 
of Science and Technology, Government of India, under PDF/2016/002148. Work of ADB is supported 
by the SERB National Post-Doctoral fellowship under this project (PDF/2016/002148). AKS would like to acknowledge 
MHRD, Govt. of India for research fellowship.
ADB and AKS also thank P. B. Pal, Anirban Biswas, Biswajit Karmakar, Rishav Roshan and Rashidul Islam
for useful help and discussions.

\end{document}